\documentclass[11pt,twoside]{article}
\usepackage{atmp}
\copyrightnotice{2003}{7}{619}{665}
\def\be{\begin{equation}}\def\ee{\end{equation}}
\def\ba{\begin{eqnarray}}\def\ea{\end{eqnarray}}
\def\cvp{\raise 2pt\hbox{,}}

\def\tr{\mathop{\rm tr}\nolimits}

\def\d{{\rm d}}

\def\uN{{\rm U}(N)}

\def\pone{{\mathbb P}^{1}}\def\O{{\cal O}}\def\F{{\cal F}}

\def\La{\Lambda}

\def\plb#1#2#3{{\it Phys.\ Lett.\ }{\bf B #1} (#2) #3}
\def\npb#1#2#3{{\it Nucl.\ Phys.\ }{\bf B #1} (#2) #3}

\def\prl#1#2#3{{\it Phys.\ Rev.\ Lett.\ }{\bf #1} (#2) #3}
\def\jhep#1#2#3{{\it J. High Energy Phys.\ }{\bf #1} (#2) #3}
\def\prd#1#2#3{{\it Phys.\ Rev.\ }{\bf D #1} (#2) #3}

\def\atmp#1#2#3{{\it Adv.\ Theor.\ Math.\ Phys.\ }{\bf #1} (#2) #3}
\def\cmp#1#2#3{{\it Comm.\ Math.\ Phys.\ }{\bf #1} (#2) #3}
\def\pr#1#2#3{{\it Phys.\ Rep.\ }{\bf #1} (#2) #3}
\def\jmp#1#2#3{{\it J.\ Math.\ Phys.\ }{\bf #1} (#2) #3}

\def\mpla#1#2#3{{\it Mod.\ Phys.\ Lett.\ }{\bf A #1} (#2) #3}

\def\jag#1#2#3{{\it J.~Algebraic Geometry }{\bf #1} (#2) #3}
\begin{document}
\title{Planar diagrams\\ and Calabi-Yau spaces}
\url{hep-th/0309151}
\author{Frank Ferrari}{\renewcommand{\thefootnote}{$\!\!\dagger$}
\address{Institut de Physique, Universit\'e de Neuch\^atel\\
rue A.-L.~Br\'eguet 1, CH-2000 Neuch\^atel, Suisse\\
and\\
Division de Physique Th\'eorique\\
CERN, CH-1211 Gen\`eve 23, Suisse}
\addressemail{frank.ferrari@unine.ch}
\markboth{\it PLANAR DIAGRAMS AND CALABI-YAU SPACES}{\it FRANK FERRARI}
\begin{abstract}
Large $N$ geometric transitions and the Dijkgraaf-Vafa conjecture
suggest a deep relationship between the sum over planar diagrams and
Calabi-Yau threefolds. We explore this correspondence in details,
explaining how to construct the Calabi-Yau for a large class of
$M$-matrix models, and how the geometry encodes the correlators. We
engineer in particular two-matrix theories with potentials $W(X,Y)$
that reduce to arbitrary functions in the commutative limit. We apply
the method to calculate all correlators $\langle\tr X^{p}\rangle$ and
$\langle\tr Y^{p}\rangle$ in models of the form
$W(X,Y)=V(X)+U(Y)-XY$ and $W(X,Y)=V(X)+YU(Y^{2})+XY^{2}$.
The solution of the latter example was not known, but when $U$ is
a constant we are able to solve the loop equations, finding a precise
match with the geometric approach. We also discuss special
geometry in multi-matrix models, and we derive an important property,
the entanglement of eigenvalues, governing the expansion around
classical vacua for which the matrices do not commute.
\end{abstract}
\cutpage
\section{Introduction}

Ever since the work of 't~Hooft on large $n$ QCD \cite{thooft},
finding methods to sum up planar diagrams has remained a central theme
in mathematical physics. Yet, even in the best studied and simplest
case of zero-dimensional integrals over hermitian $n\times n$ matrices,
only very few techniques
are available \cite{kaza1}. The usual approach is to solve a set of
saddle point equations by making an ansatz for the analytic structure
of the solution \cite{BIPZ}. The saddle point equations are themselves
derived after reducing the number of degrees of freedom from $\sim n^{2}$
to $\sim n$ by integrating over angular variables, which can necessitate
sophisticated methods \cite{meh,IZint,cha} that apply only in special cases.
Another approach is to write down Schwinger-Dyson equations, called
loop equations in this context \cite{loopk}. This method is very
general and does not use assumptions on the analytic structure of the
solution. However, it is usually very difficult to find a finite set
of equations that closes under the correlators one wish to calculate.
This scarcity of tools and of solvable examples makes the search for
alternative strategies particularly useful. 

It is well-known that the solution of the one-matrix model can be
expressed in terms of a complex algebraic curve (see for example
\cite{revmat} or the appendix of \cite{fer1} for reviews). The idea we
wish to pursue in the following is that for multi-matrix models it may
be useful to consider higher dimensional Calabi-Yau spaces. This is
suggested by conjectures by Vafa and collaborators on the strongly
coupled dynamics of four dimensional ${\cal N}=1$ supersymmetric gauge
theories \cite{GV,CIV,CKV,DV}. We will engineer multi-matrix models
with potentials of the form
\be\label{potential} W(X_{1},\ldots,X_{M})={1\over 2i\pi}\oint_{C_{0}}
z^{-M-1} E\Bigl( z,\sum_{i=1}^{M}X_{i}z^{i-1}\Bigr) \, \d z
\, ,\end{equation}
where $E(z,w)=\sum_{i=-\infty}^{+\infty} E_{i}(w)\,
z^{i}$ can be expanded in terms of entire functions $E_{i}$ and 
$C_{0}$ is a small contour encircling $z=0$. 
This is a large and interesting class of models.
In the one- or two-matrix cases, the commutative limit of
(\ref{potential}) can be an arbitrary function.

The main body of the paper is divided into four parts. In Section 2
and Appendix A, we discuss special geometry relations for multi-matrix
integrals. Special geometry is automatically built in the Calabi-Yau
geometry, and it is important to have a satisfying understanding from
the matrix model perspective as well. Deriving the relations, we
uncover new interesting properties of the saddle point equations and
their solutions (eigenvalue entanglement) around vacua where the
matrices do not commute. These developments in standard matrix model
technology are not strictly required to understand the geometric
approach, that we describe in details in Sections 3 and 4. The method
is applied in particular to compute the resolvents of both matrices
$X$ and $Y$ for the standard \cite{2mm} two-matrix model $W(X,Y)= V(X)
+ U(Y) - XY$ as well as for the model $W(X,Y) = XY^{2} + V(X) +
YU(Y^{2})$, for arbitrary polynomials $V$ and $U$. In Section 5, we
use the loop equation technique to solve for all correlators
$\langle\tr X^{p}Y^{q}\rangle$ in the latter example when $U$ is a
constant, finding a perfect match with the geometric
approach. Finally, we discuss open directions of research in Section
6.

\section{The geometry of matrix models}
\setcounter{equation}{0}

The existence of special geometry relations constraining the sum over
planar diagrams is a basic ingredient of the geometric approach to
matrix models. This is an essentially unexplored subject in the case of
multi-matrix integrals, and thus we provide a rather detailed, albeit
incomplete, discussion in the present Section and in Appendix A.

A matrix model with $M$ matrices has several one-matrix model effective
descriptions obtained by integrating out all but one matrix, for 
example
\be\label{veffgen}
e^{-{n^{2}\over S} V_{{\rm eff}}(X)} = \int\prod_{i=2}^{M}\d X_{i}\,
e^{-{n\over S}\tr W(X,X_{2},\ldots ,X_{M})} .\end{equation}
By analogy with the ordinary one-matrix model with polynomial
potential, it is natural to suspect that special geometry relations
could be derived for the multi-matrix models by working with the
effective descriptions. The difficulty is that the effective potential
$V_{\rm eff}$ can have a non-trivial analytic structure. The main
subtleties come from the classical solutions for which the matrices do
not commute. By studying the properties of the effective potentials,
we derive a remarkable property of the eigenvalue distributions
governing the expansion around such solutions. Full details are
provided in Appendix A for two models that display some generic
features.

The various sets of special geometry relations, derived for each
well-behaved one-matrix effective description, must be equivalent. In
the geometric approach, this non-trivial constraint is automatically
satisfied by expressing the solution in terms of integrals over a
higher dimensional complex variety equipped with prefered coordinates
associated with the various matrices.

\subsection{Definitions}

We consider a model with $M$ hermitian $n\times n$ matrices 
$X_{k}$ and polynomial potential $W(X_{1},\ldots X_{M})$. The 
classical equations of motion 
\be\label{cleq} {\cal A}:\ \tr\d W(X_{1},\ldots ,X_{M}) = 0 \end{equation}
define an algebra $\cal A$ with a set of $D_{K}$-dimensional
irreducible representations $R_{K}$. In the generic case there are no
moduli and the index $K$ is discrete.

A particular solution to (\ref{cleq}) of the form
\be\label{vac}R=\oplus n_{K}R_{K}\, , 
\quad\sum_{K}n_{K}D_{K}=n\, ,\end{equation}
is called a classical vacuum. 
The partition function $\F(R)$ is defined by the formula
\be \label{def}
e^{n^{2}\F(R)/S^{2}} = \int_{R-{\rm planar}}\d X_{1}\cdots\d X_{M}\,
e^{-{n\over S}\tr W(X_{1},\ldots,X_{M})}\, ,\end{equation}
where the subscript ``$R$-planar'' in (\ref{def}) means that we sum up
only the contributions from the planar diagrams in the perturbative
Feynman expansion around the classical solution $R$. The partition
function depends of the integers $n_{K}$ in (\ref{vac}), or more
conveniently on the variables
\be\label{Skdef} S_{K} = {n_{K}\over n}\, S\end{equation}
that satisfy the identity
\be\label{Sconst}\sum_{K}D_{K}S_{K} = S\, .\end{equation}
In general, the perturbative expansion defining $\F$
involves ghost matrices as
explained in \cite{pert}. The coupling $S$ plays the r\^ole of
$\hbar$. An important result is that the sum over planar diagrams is
always absolutely convergent for small enough $S$ \cite{thooftcv}. For
that reason the convergence properties of the integral $\int\d
X_{1}\cdots\d X_{M}\exp (-n\tr W/S)$ are irrelevant. It can always be
made convergent without loss of generality by performing an analytic
continuation and/or turning on some couplings. If the integral is
convergent, then its strict large $n$ limit yields $\F(R_{\rm min})$
for the representation $R_{\rm min}$ that minimizes $W$ (assuming that
it is unique). A more useful statement is that for general $R$,
$\F(R)$ is the large $n$ limit of the formal perturbative series
around the classical solution $R$. In this limit the variables 
$S_{K}$ defined in (\ref{Skdef}) become continuous.

Correlators are defined by
\be \langle{\cal O}\rangle_{R} = e^{-n^{2}\F(R)/S^{2}}
\int_{R-{\rm planar}}\d X_{1}\cdots\d X_{M}\, {\cal O}\,
e^{-{n\over S}\tr W(X_{1},\ldots,X_{M})}\, .\end{equation}
We will not systematically indicate the dependence in $R$.
Basic objects that we want to compute are the resolvents of the 
matrices $X_{m}$ defined on the complex plane by
\be\label{defres} g^{X_{m}}(x) = S\,\Bigl\langle {\tr\over n}\, {1\over 
x-X_{m}}\Bigr\rangle\, .\end{equation}
The resolvents are the generating functions of the correlators
$\langle\tr X_{m}^{p}\rangle$ for arbitrary positive integers $p$.
They are analytic functions with square root branch cuts. The physical
sheet is defined by the asymptotics
\be\label{asy} g^{X_{m}}(x) \mathrel{\mathop{\kern 0pt\sim}
\limits_{x\rightarrow\infty}^{}} {S\over x}\,\cdotp\end{equation}
The densities of eigenvalues $\rho^{X_{m}}$ of the matrices $X_{m}$ 
are read off from discontinuities of 
the resolvents across branch cuts on the physical sheet,
\be\label{densfor} \rho^{X_{m}}(x) = {i\over 2\pi S}
\left(\strut g^{X_{m}} (x +i\epsilon) - 
g^{X_{m}} (x -i\epsilon)\right)\, ,\end{equation}
and we have
\be\label{resfor} g^{X_{m}}(x) = S
\int_{-\infty}^{+\infty}{\rho^{X_{m}}(z)\,\d z\over x-z}\,\cdotp\end{equation}
\subsection{Special geometry}
\subsubsection{Generalities}

We work with the one-matrix model whose potential is
defined by (\ref{veffgen}). For a general vacuum (\ref{vac}), the
eigenvalues of $X$ form cuts $I_{K,k}^{X}$, $1\leq k\leq D_{K}$,
that generically do not overlap. The number of eigenvalues in the
cuts, or equivalently the variables $S_{K}$ defined in (\ref{Skdef}),
are expressed in terms of period integrals,
\be\label{sgea}
S_{K} = {1\over 2i\pi}\oint_{\alpha^{X}_{K,k}} g^{X}(x)\,\d x \, ,\end{equation}
where the contours $\alpha^{X}_{K,k}$ encircle the cuts $I^{X}_{K,k}$ on
the physical sheet for $g^{X}$ (see Figure 1 for the definition of
various contours). The formula (\ref{sgea}) could be written
in many different-looking but equivalent ways, using other
matrices $X_{i}$ of the original multi-matrix model.
\begin{figure}
\centerline{\epsfig{file=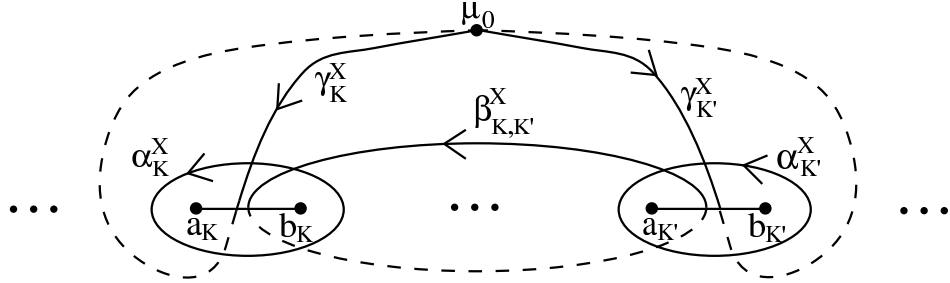,width=12.5cm}}
\caption{Definition of the contours
$\alpha^{X}_{K}$, $\beta^{X}_{K,K'}$ and $\gamma^{X}_{K}$ used in the
main text (similar contours like $\tilde\alpha^{X}_{K}$ or
$\alpha^{Y}_{K}$ are also used). The eigenvalues that classically sit
in the representations $R_{K}$ and $R_{K'}$ are spread along the cuts
$I^{X}_{K}=[a_{K},b_{K}]$ and $I^{X}_{K'}=[a_{K'},b_{K'}]$
respectively.
\label{fig1}}
\end{figure}
Taking into account the constraints (\ref{sgea}), the eigenvalues
adjust to make the partition function
\be\label{gpartf}
\F = - S V_{\rm eff}(x_{1},\ldots ,x_{n}) + {S^{2}\over n^{2}}
\sum_{i\not = j}\ln |x_{i}-x_{j}|\end{equation}
extremal. The corresponding saddle point equations
\be\label{gsp}
- n {\partial V_{\rm eff}\over\partial x_{i}}+
{2 S\over n}\sum_{j\not = i}{1\over x_{i}-x_{j}}=0\end{equation}
are interpreted as vanishing force conditions. The total force on a 
test eigenvalue on the complex plane
\be\label{totf} f(x) = f_{\rm b}(x) + 2g^{X}(x)\end{equation}
is the sum of a background force
\be\label{forcegdef} f_{\rm b}(x)= - n {\partial V_{\rm eff}\over\partial 
x_{i}}\lower 3pt\hbox{$\Bigr|_{x_{i}=x}$},\end{equation}
and of the Coulomb term $2g^{X}(x)$. Since varying $S_{K}$ amounts to
moving eigenvalues from one cut to another \cite{DV}, we expect that
$\partial\F/\partial S_{K}$ will be expressed in terms of countour
integrals of the differential $f\d x$. If the term $f_{\rm b}\d x$
does not contribute to the integrals, as in the ordinary one-matrix
model for which it is an exact differential $-\d W$, the special
geometry relations associated with (\ref{sgea}) straightforwardly
follow, for example
\be\label{sgeb}
{\partial\F\over\partial S_{K}} =
\oint_{\gamma^{X}_{K}} g^{X}(x)\,\d x + {\rm counterterm}\, ,\end{equation}
where the non-compact $\gamma$-cycles are defined in Figure 1. However, in
general, $f_{\rm b}\d x$ do contribute, and a more detailed discussion
is needed.
\subsubsection{The commutative case}
There is a class of models for which the classical equations of motion
imply that the matrices commute with each other. The non-commutative
structure can then, in some sense, be neglected. This is also true
more generally for vacua built from one dimensional representations
only. A typical example is the much-studied two-matrix model with
potential
\be\label{2mmpot} W(X,Y) = V(X) + U(Y) - XY\, ,\end{equation}
where $V$ and $U$ are polynomials of degree $d_{X}+1$ and
$d_{Y}+1$ respectively. This model was used for example in the study
of two dimensional gravity coupled to $c<1$ matter \cite{revmat}. The
classical equations of motion
\be\label{2mmcl}{\cal A}:\
X = U'(Y) = Q(Y)\, ,\quad Y = V'(X) = P(X)\, ,\end{equation}
imply that $[X,Y]=0$, and thus the only irreducible representations 
are the $d_{X}d_{Y}$ one dimensional representations given by
$X=x I$ and $Y=yI$ with
\be\label{2mmirrep} x = Q \circ P(x)\, ,\ y = P(x)\quad
\Longleftrightarrow\quad y = P\circ Q(y)\, ,\ x = Q(y)\, .\end{equation}
The effective potential for $X$ is the sum of a classical part $(\tr
V)/n$ and of a quantum term proportional to the partition function
of the one-matrix model in an external field. For cubic $U$, this
partition function is related to the Kontsevich integral for
two-dimensional topological gravity \cite{kont}, and was studied in
\cite{1mmext}. In general, a qualitative picture of the analytic
structure of the background force can be obtained by assuming that $S$
is very small. In this limit the matrix $Y$ can be integrated out
classically and we get
\be\label{fcl2mm} f_{\rm b,\, cl}(x) = -P(x) + Q^{-1}(x)\, .\end{equation}
The functional inverse $Q^{-1}$ of the polynomial $Q$ is a
multi-valued function with $d_{Y}$ sheets. There are $d_{Y}-1$
branching points at $x={Q'}^{-1}(0)$. The analytic structure of $f_{\rm
b}$ at finite $S$ is a perturbation of the analytic structure of
$f_{\rm b,\, cl}$, with the same number of sheets and of branching
points. This can be explicitly checked when $U$ is cubic by using
the results of \cite{1mmext}, and no other singularities are expected.
In particular, $f_{\rm b}$ does not have branch cuts on the support of
$\rho^{X}$, and thus does not contribute to contour integrals over
$\alpha$-, $\beta$- or $\gamma$-cycles. The relations (\ref{sgeb}) are
thus valid. Moreover, the saddle point equation (\ref{gsp}) has a
simple $n\rightarrow\infty$ limit,
\be\label{ninf1} f_{\rm b}(x) + g^{X}(x+i\epsilon) + 
g^{X}(x-i\epsilon) = 0 \quad {\rm for}\ x\in {\rm 
Support}[\rho^{X}]\, ,\end{equation}
showing that $g^{X}$ is a $(d_{Y}+1)$-sheeted analytic
function (the physical sheet and the $d_{Y}$ additional sheets of
$f_{\rm b}$). This is consistent with the fact that $g^{X}$ is an
algebraic function of degree $d_{Y}+1$ (see \cite{2mm}
and Section 4). A symmetric discussion applies to the effective
description in terms of the matrix $Y$.
\subsubsection{The non-commutative case}
Things are much more subtle and interesting when higher dimensional
representations are present. The na\"\i ve classical limit of a saddle
point equation like (\ref{ninf1}) would yield the one-dimensional
representations only. The emergence of the higher dimensional
representations, which is a classical effect in the description in
terms of the $M$ matrices $X_{i}$, must come out as a quantum effect
in the effective description in terms of the single matrix $X$. This
means that some of the quantum corrections in $f_{\rm b}$ are singular
and can have a finite $S\rightarrow 0$ limit that modifies the na\"\i
ve classical limit. A related crucial fact is that, even though the
equations (\ref{gsp}) are expressed in terms of the eigenvalues of $X$
only, they must know about correlations between $X$ and the other
matrices that do not commute with $X$. This implies that in the
continuum limit, (\ref{gsp}) cannot be expressed in terms of
$\rho^{X}$ only. We will argue that the density $\rho^{X}$ is still
unambiguously determined, because there is a constraint, the
entanglement of eigenvalues, that relates $\rho^{X}$ on the various
intervals $I^{X}_{K,k}$ associated with a given representation
$R_{K}$. Intuitively, this constraint comes from the ${\rm U}(n)$
gauge symmetry of the original matrix model.

Those facts are very general but are best illustrated on
an example. Let us consider the model with potential
\be\label{ldef} W(X,Y) = XY^{2} + \alpha Y + V(X)\end{equation}
for an arbitrary polynomial $V$ of degree $d+1$. It is useful to 
separate $V'(X)$ into an even and an odd part,
\be\label{F1F2def} -V'(X) = F_{1}(X^{2}) + X F_{2}(X^{2})\, .\end{equation}
The classical equations of motion
\be\label{lcleq}{\cal A}:\ \{X,Y\} = -\alpha\, ,\quad 
Y^{2}=F_{1}(X^{2}) + X F_{2}(X^{2})\, ,\end{equation}
admit $d+2$ irreducible representations $R_{K}$ of dimension one,
$X=xI$ and $Y=yI$ with
\be\label{l1rep} 2 x y = -\alpha\, ,\quad y^{2}= - V'(x)\, ,\end{equation}
and $[(d-1)/2]$ irreducible representations $\tilde R_{K}$ of
dimension two for which, in terms of Pauli matrices, $X=x\sigma^{3}$
and $Y=y_{1}\sigma^{1} +  y_{3}\sigma^{3}$ with
\be\label{l2rep} 2 x y_{3} = -\alpha\, ,\quad y^{2}= y_{1}^{2}
+ y_{3}^{2} = F_{1}(x^{2})\, ,\quad F_{2}(x^{2}) = 0\, .\end{equation}
If only the one dimensional representations are considered, then the
saddle point equations for the matrix $X$ have standard features and
can be solved by the usual tricks. This was done for $\alpha=0$ and with
only one cut that do not intersect the origin in \cite{EZJ}. We are
interested in the general case. The effective potential for $X$, 
the background force and the saddle point equations
can be calculated exactly,
\ba &&\hskip -1.5cm V_{\rm eff}(x_{1},\ldots x_{n}) = {1\over 
n}\sum_{i} \Bigl( V(x_{i})-{\alpha^{2}\over 
4x_{i}}\Bigr) + {S\over 2n^{2}}\sum_{i\not =j}\ln |x_{i}+x_{j}|\, 
,\label{veffX}\\
&& \hskip -1.5cm f_{\rm b}(x) = -V'(x) - {\alpha^{2}\over 4 x^{2}}-
{S\over n}\sum_{j}{1\over x+x_{j}} = -V'(x) - {\alpha^{2}\over 4 
x^{2}} + g^{X}(-x)\, ,\label{laubf}\\
&& \hskip -1.5cm -V'(x_{i}) - {\alpha^{2}\over 4 x_{i}^{2}}-
{S\over n}\sum_{j}{1\over x_{i}+x_{j}} +
{2 S\over n}\sum_{j\not = i}{1\over x_{i}-x_{j}} = 0\, .\label{sadL}\ea
It is important to assume that $\alpha\not =0$, because otherwise the
vacua $X=0$, $Y=\pm \sqrt{F_{1}(0)}\,\sigma^{1}$, for which the
$\mathbb Z_{2}$ symmetry $Y\rightarrow -Y$ is broken, can no longer be
described after the exact integration over $Y$. If needed, the limit
$\alpha\rightarrow 0$ can be taken on the final formulas. The
singularities in (\ref{veffX}) and (\ref{laubf}) can be understood as
follows. Classically, integrating out $Y$ amounts to solving the
equation
\be\label{outY}\{ X,Y\} + \alpha = 0\end{equation}
for a given matrix $X$, and this implies that
$[X^{2},Y]=0$. Generically (\ref{outY}) has a unique solution $Y =
-\alpha X^{-1}/2$, which is singular if $X$ is not invertible,
yielding poles at $x_{i}=0$. More interestingly, if two eigenvalues of
$X$ are such that $x_{i}=-x_{j}$, then a different class of solutions
to (\ref{outY}) is possible, corresponding to the two-dimensional
representations. In that case, the $S\rightarrow 0$ limit of the term
$S\sum_{j}1/(x_{i}+x_{j})$ in (\ref{sadL}) is non-zero and compensates
for the non-zero even background force in (\ref{l2rep}).

A typical configuration with a two-dimensional representation $\tilde
R_{K}$, corresponding to a solution $x^{2}=x^{2}_{K}$ of
(\ref{l2rep}), is depicted in Figure 2.
\begin{figure}
\centerline{\epsfig{file=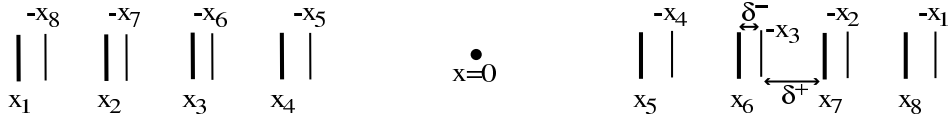,width=12.5cm}}
\caption{Eigenvalues filling a two dimensional representation $\tilde 
R_{K}$ for $n_{K}=4$ (thick lines) and their images with 
respect to $x=0$  (thin lines). When $\delta^{+}\not =\delta^{-}$ the 
eigenvalues feel a quantum even force that compensates for the 
non-zero classical force $F_{1}-\alpha^{2}/(4x^{2})$ in (\ref{l2rep}).
\label{fig2}}
\end{figure}
We put $n_{K}$ eigenvalues $x_{1},\ldots ,x_{n_{K}}$ at $x=-x_{(K)}$
and $n_{K}$ eigenvalues $x_{n_{K}+1},\ldots ,x_{2n_{K}}$ at
$x=x_{(K)}$. Turning on $S$, the eigenvalues fill intervals $\tilde
I^{X}_{K} = [\tilde a_{K},\tilde b_{K}]$ and $\tilde I^{'X}_{K} =
[\tilde a'_{K},\tilde b'_{K}]$ around $x_{(K)}$ and $-x_{(K)}$
respectively. The attractive force between eigenvalues and their images,
described by the term $S\sum_{j} 1/(x_{i}+x_{j})$, induces strong
correlations between the eigenvalues around $+x_{K}$ and the
eigenvalues around $-x_{K}$. If we choose the labels such that
$x_{1}<x_{2}<\cdots <x_{2n_{K}}$, classically $x_{1}=\cdots
=x_{n_{K}}=-x_{(K)}$ and $x_{n_{K}+1}=\cdots=x_{2n_{K}}=x_{(K)}$.
Quantum mechanically, the only stable, finite energy configurations are
then such that $-x_{2n_{K}}<x_{1}$ and $x_{i}<-x_{2n_{K}-i}<x_{i+1}$
for $1\leq i\leq n_{K}-1$, or $x_{i}<-x_{2n_{K}-i+1}<x_{i+1}$ for
$1\leq i\leq n_{K}-1$ and $x_{n_{K}}<-x_{n_{K}+1}$. This entanglement
of eigenvalues can be proven as follows. If more than one image
eigenvalue $-x_{j}$, $j> n_{K}$, were in a given interval
$]x_{i},x_{i+1}[$, $i\leq n_{K}-1$, then, because of the attractive
force between eigenvalues and image eigenvalues, the system would
collapse to a singular configuration with divergent energy. If there
were some intervals $]x_{i},x_{i+1}[$ without image eigenvalues then,
due to the repulsive force, the eigenvalues that are not entangled
would not converge to the equilibrium positions $x_{K}$ or $-x_{K}$ for
the two-dimensional representations $\tilde R_{K}$ when $S\rightarrow
0$.

A direct consequence of the entanglement is that, in the continuum
large $n$ limit, the intervals $\tilde I_{K}$ and $\tilde
I'_{K}$ are symmetric, $\tilde b'_{K} = -\tilde a_{K}$ and $\tilde
a'_{K}=-\tilde b_{K}$, and
\be\label{evenrho} \rho^{X}(x) = \rho^{X}(-x)\quad {\rm for}\ x\in 
\tilde I^{X}_{K}\, .\end{equation}
This shows that the background force (\ref{laubf}) has branch cuts
that coincide with those of $g^{X}$, and in particular will contribute
to contour integrals. Moreover, the continuum limit of the saddle
point equation (\ref{gsp}, \ref{sadL}) must be studied with care. One might want
to replace $f_{\rm b}(x)$ by $(f_{\rm b}(x+i\epsilon) + f_{\rm
b}(x-i\epsilon))/2$, but this is not correct. As explained in Appendix
A, Section 1, if $\delta^{+} (x)$ and $\delta^{-} (x)$ are the even
functions defined in Figure 2, then the continuum limit is
\ba &&\hskip -1.2cm {1\over 2}\left(\strut
f_{\rm b}(x+i\epsilon) + f_{\rm b}(x-i\epsilon)\right) + g^{X}
(x+i\epsilon) + g^{X}(x-i\epsilon) = \nonumber\\&&\hskip 2.5cm
S\rho^{X}(-x)\ln{\delta^{-}(x)\over\delta^{+}(x)}\, \cvp
\quad {\rm for}\ x\in {\rm Support}[\rho^{X}]\, .\label{spnc}\ea
The even ``quantum'' force on the right hand side of (\ref{spnc}) can
compensate for the even classical force
$F_{1}(x^{2})-\alpha^{2}/(4x^{2})$ in the $S\rightarrow 0$ limit.
Consequently, there can exist classical equilibrium configurations for
which only the odd part $xF_{2}(x^{2})$ of the classical force
vanishes. Those configurations correspond precisely to the two
dimensional representations (\ref{l2rep}). By analysing
(\ref{evenrho}) and (\ref{spnc}), we further show in Appendix A that
$g^{X}$ is generically three-sheeted. It satisfies a degree three
algebraic equation that we compute by two different methods in Sections 
4 and 5.\vfill\eject
\subsubsection{Lesson}
For higher dimensional representations, we have seen that the
continuum saddle point equation involves additional unknown functions
and is supplemented by conditions coming from the entanglement of
eigenvalues. Once this is understood, one can proceed and try to
derive the special geometry relations. This is done is Appendix A for
the model (\ref{ldef}), both in the description in terms of $X$
discussed above and in the description in terms of $Y$. The same
analysis applies straightforwardly in a large class of models,
including the famous ${\cal N}=1^{*}$ theory, in which case
an ansatz corresponding to eigenvalue entanglement was first made in
\cite{N1star}.

In general, the effective potential, as a function of the eigenvalues
of a given matrix $X$, has singularities when the eigenvalues satisfy
some conditions corresponding to the emergence of representations for
which $X$ does not commute with all the other matrices. In our case
the condition was $x_{i}=-x_{j}$, but other type of conditions, for
example $x_{i}-x_{j}\in\mathbb Z$, are possible. The logarithmic
singularities that we found in (\ref{veffX}) are singled out by the
fact that $V_{\rm eff}$ is a well-defined functional of $\rho^{X}$ in
the continuum limit (which is necessary for the idea of a master
field, and the corresponding factorization of correlation functions,
to apply), unlike its derivatives, that know about the correlations
between $X$ and the other matrices. The generated force must be
attractive in order to stabilize new higher dimensional solutions in
the classical limit. These features presumably imply eigenvalue
entanglement and special geometry in a large class of models.

\subsection{Solving the constraints from special geometry}

A natural way to implement special geometry is to express the 
solution in terms of a Calabi-Yau manifold with nowhere vanishing
holomorphic top-form $\Omega$,
\be\label{solCY1}
S_{K}={1\over 2i\pi}\oint_{A_{K}}\!\! \Omega\, ,\quad
{\partial\F\over\partial S_{K}}= 
\oint_{C_{K}}\!\! \Omega + {\rm counterterm}\, .\end{equation}
The $A_{K}$ and $C_{K}$ are holomorphic spheres in one-to-one
correspondence with the $\alpha^{X_{m}}_{K}$ and $\gamma^{X_{m}}_{K}$
cycles. A very important additional property is that, at least in a
large class of examples, there are privileged coordinates associated
with the matrices, or more precisely with the generators of the center
of the algebra $\cal A$. Integrating over the coordinates in different
orders in (\ref{solCY1}), we can obtain the description in terms of
the various matrices. We will argue in Section 4 that this property
makes possible the computation of the resolvents.

\section{Matrix models and Calabi-Yau threefolds}
\setcounter{equation}{0}

Let us consider type IIB string theory one some non-compact Calabi-Yau
threefold. Let us assume that there is a $\pone$ with normal bundle
$\cal N$ and quantum volume $V$ in the geometry. By wrapping $N$ D5
branes on the $\pone$, we engineer a four dimensional ${\cal N}=1$
supersymmetric gauge theory, with bare Yang-Mills coupling constant
$g_{\mu_{UV}}^{2}\sim 1/V$ and gauge group $\uN$. The deformation
space of the $\pone$, or equivalently the space of fluctuations of the
brane, is generated locally by the holomorphic sections of the normal
bundle $\cal N$. Physically, this means that if $\cal N$ has $h$
linearly independent holomorphic sections, the gauge theory is coupled
to $h$ scalar superfields in the adjoint representation. For a generic
geometry, there is an obstruction to the deformation space, which in 
favorable cases can be interpreted physically in terms of a 
superpotential $W$.

If $h\leq 2$, the resulting gauge theory is asymptotically free and
dynamically generates a mass gap $\La$. The renormalized Yang-Mills
coupling $g_{\mu}^{2}$ at scale $\mu$ grows in the IR, and becomes
infinite at some scale $\mu_{\rm c}\sim\La$. Beyond that scale the
fundamental gauge fields strongly fluctuate and are no longer
appropriate degrees of freedom. From the type IIB perspective the RG
flow corresponds to the shrinking of the $\pone$ to zero size. At
scale $\mu=\mu_{\rm c}$, the $\pone$ vanishes and the original smooth
Calabi-Yau is replaced by a singular geometry. What happens for $\mu
<\mu_{\rm c}$ has not been derived from first principles, but there
are remarkable conjectures that partially address the problem. On the
gauge theory side, it is believed that good degrees of freedom are
given by the so-called glueball superfields $S_{i}$, and that the
exact quantum superpotential $W_{\rm glueballs}$ for the $S_{i}$ can
be calculated from the partition function of an associated $h$-matrix
model whose potential is equal to the superpotential of the gauge
theory \cite{DV}. On the string theory side, it is believed that the
singular Calabi-Yau is deformed to a smooth Calabi-Yau. The branes and
the $\pone$ disappear and are replaced by three-form flux
through three-spheres. The non-zero flux generates a superpotential
$W_{\rm flux}$ that can be computed from special geometry \cite{flux}.
The consistency between the conjectures,
\be\label{consis}W_{\rm glueballs}=W_{\rm flux}\, ,\end{equation}
implies a highly non-trivial relationship between the partition function
of the matrix model and the Calabi-Yau geometry, whose form coincide
precisely with (\ref{solCY1}). This relationship is indicated by a
question mark in Figure 3. 

\begin{figure}
\centerline{\epsfig{file=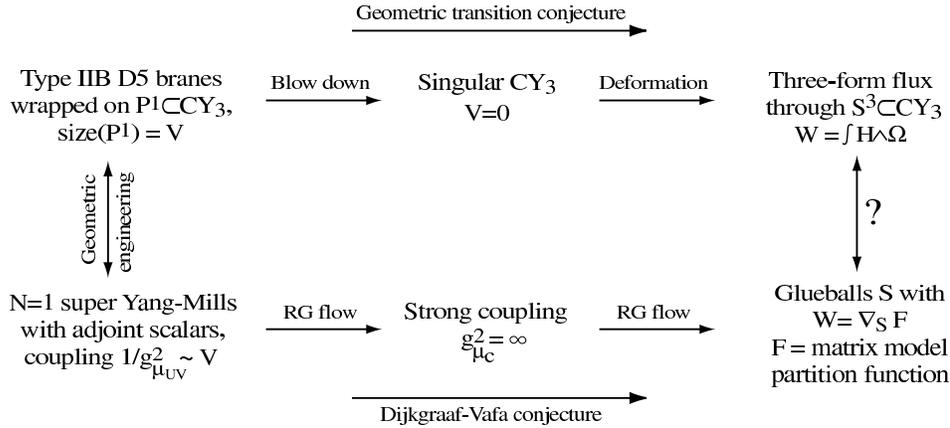,width=12.5cm}
}
\caption{Chain of conjectured dualities suggesting a relation between 
Calabi-Yau spaces and matrix models. The question mark emphasizes the 
link that we want to explore. See details in the main text.
\label{fig}}
\end{figure}

The great advantage of this gauge theory/string theory set-up is
that we have a precise recipe, explained below, to
construct the relevant Calabi-Yau spaces. On the other hand, it is
important to emphasize that there is no general proof of the validity
of the approach. For example, the full relationship between ${\cal
N}=1$ gauge theories and matrix models can be proven only in special
cases and under some assumptions \cite{ferp}.
A perturbative argument can also be given \cite{pproof}, but it does
not apply to the objects that are relevant for us. In particular,
there is no understanding of ${\cal N}=1$ special geometry \cite{LM}
in gauge theory from first principles.

As is clear from Figure 3, there are three geometries that play a
r\^ole, and that we discuss in turn in the following. The first
geometry is the original Calabi-Yau space $\hat{\cal M}$ containing
the $\pone$s with normal bunble $\cal N$ on which we wrap the D5
branes. This space $\hat{\cal M}$ is called the resolved Calabi-Yau,
because it is a small resolution of the singular Calabi-Yau ${\cal
M}_{0}$ obtained from $\hat{\cal M}$ by shrinking the $\pone$ (the
resolution is small because the singular points are replaced by
curves). Precisely, there exists a blow down map $\pi :\hat{\cal
M}\rightarrow {\cal M}_{0}$ that locally induces a birational
isomorphism between $\hat{\cal M}\backslash\{\pone\}$ and ${\cal
M}_{0}\backslash\{ p\}$ where $p$ is a singularity and
$\pi^{-1}(p)=\pone$. Finally there is the smooth deformed space $\cal
M$ obtained by perturbing the algebraic equation defining ${\cal
M}_{0}$. All we need to know about $\cal M$ is this algebraic
equation, that encodes the complex structure, because the metric datas
are irrelevant in integrals like (\ref{solCY1}).\vfill

\subsection{The resolved geometry}
\subsubsection{The transition functions}
Let us describe $\hat{\cal M}$ by two coordinate patches
$(z,w_{1},w_{2})$ and $(z',w'_{1},w'_{2})$, where $z$ and $z'=1/z$ are
the stereographic coordinates on the two-spheres, and consider the
following transition functions
\be \label{geo1}
z'=1/z\, ,\quad w'_{1}=z^{-n}w_{1}\, ,\quad w'_{2}=z^{-m}w_{2}\, .\end{equation}
If $n\geq 0$ and $m<0$ (other cases could be discussed as well, but 
are irrelevant for our purposes), the geometry (\ref{geo1}) has a 
$(n+1)$-dimensional continuous family of $\pone$s that sit at
\be\label{sec1}
w_{1}(z) = \sum_{i=1}^{n+1} x_{i} z^{i-1}\, ,\quad  w_{2}(z)=0\, .\end{equation}
The normal bundle to the $\pone$s is ${\cal N}=\O(n)\oplus\O(m)$ by
construction. The functions $w_{1}(z)$ and $w_{2}(z)$ in (\ref{sec1})
are characterized by the fact that they must define globally
holomorphic sections of $\cal N$. This is indeed the case because in 
the other coordinate patch, (\ref{geo1}) yields
\be\label{sec2} w'_{1}(z') = \sum_{i=1}^{n+1} x_{i} {z'}^{n-i+1}\, ,\quad
w'_{2}(z')=0\, .\end{equation}
The parameters $x_{i}$, $1\leq i\leq n+1$, span the versal deformation
space of the $\pone$s. A more interesting geometry is obtained by
perturbing the transition functions given by (\ref{geo1}). The
perturbation is described in terms of a geometric potential. The
potential is a function $E(z,w)$ of two complex variables that can be
Laurent expanded in terms of entire functions $E_{i}$,
\be\label{Ugexp}E(z,w) = \sum_{i=-\infty}^{+\infty} 
E_{i}(w) z^{i}\, .\end{equation}
The most general geometry we will consider is given by
\be \label{geogen}
z'=1/z\, ,\quad w'_{1}=z^{-n}w_{1}\, ,\quad w'_{2}=z^{-m}w_{2} + 
\partial_{w}E(z,w_{1})\, .\end{equation}
Since the relation between $w_{1}$ and $w'_{1}$ is unchanged with
respect to (\ref{geo1}), the most general holomorphic section
$(w_{1}(z),w_{2}(z))$ of $\cal N$ is still such that
\be\label{w1exp} w_{1}(z) = \sum_{i=1}^{n+1} x_{i} 
z^{i-1}\Longleftrightarrow w'_{1}(z') = \sum_{i=1}^{n+1} x_{i} 
{z'}^{n-i+1}\, .\end{equation}
On the other hand, $w_{2}(z)$ must now be adjusted non-trivially to
insure that $w'_{2}(z')$ is holomorphic. Because of the term
$z^{-m}w_{2}={z'}^{-|m|}w_{2}$ in (\ref{geogen}), one can always choose
$w_{2}(z)$ to cancel all the poles in ${z'}^{-j}$ for $j\ge |m|$, but
there remains $|m|-1$ singular terms that cancel only if the
parameters $x_{i}$ in (\ref{w1exp}) satisfy $|m|-1$ constraints. Thus
we find that the versal deformation space of $\pone$ is spanned by
$n+1$ parameters satisfying $|m|-1$ constraints (the same result could
be straightforwardly obtained for the most general perturbation of
(\ref{geo1}), see \cite{geoalg}). Since the tangent bundle to $\pone$
is ${\cal O}(2)$, the Calabi-Yau condition of vanishing first Chern
class for the total space is
\be\label{CYcond} M=n+1=-m-1\end{equation}
and is thus equivalent to the equality of the number of parameters and
the number of constraints.
\subsubsection{The superpotential}
The special form (\ref{geogen}) of the perturbation is particularly
interesting because in this case the constraints are integrable and
are equivalent to the extremization $\d W=0$ of a superpotential
$W(x_{1},\ldots ,x_{n+1})$. This was proven in \cite{katzv} in the
case where $E(z,w)$ is regular at $z=0$, and the same argument
could be straightforwardly generalized to (\ref{Ugexp}). We can 
actually derive an explicit formula for $W$ that makes this property 
manifest. By using (\ref{geogen}), (\ref{w1exp}) and (\ref{CYcond}) 
we obtain
\be w'_{2}(z') = z^{M+1}w_{2}(z) + \partial_{w}
E\Bigl(z,\sum_{i=1}^{M}x_{i}z^{i-1}\Bigr).\end{equation}
Let us define the coutour $C_{z}$ to encircle both the points $u=0$ 
and $u=z$. If we choose in a manifestly $z$-holomorphic way $w_{2}$ to be
\be\label{w2gen} w_{2}(z) = -{1\over 
2i\pi}\oint_{C_{z}}{\partial_{w}E
\left(\strut u,\smash{\sum_{i=1}^{M}}x_{i}u^{i-1}\right)\d u\over u^{M+1} 
(u-z)}\,\cvp\end{equation}
and impose the $M$ integrable constraints
\be\label{cgen}
{\partial W\over\partial x_{i}} = {1\over 2i\pi}\oint_{C_{0}}\d z\, 
z^{i-M-2}\partial_{w}
E\Bigl(z,\sum_{j=1}^{M}x_{j}z^{j-1}\Bigr) = 0\, ,\quad 1\leq i\leq 
M\, ,\end{equation}
then we get a manifestly $z'$-holomorphic $w'_{2}$,
\be\label{wp2gen} w'_{2}(z') = {1\over 
2i\pi}\oint_{C_{z'}}{\partial_{w}E
\left(\strut 1/u,\smash{\sum_{i=1}^{M}}x_{i}u^{1-i}\right)\d u\over  
u-z'}\,\cdotp\end{equation}
The superpotential is
\be\label{potcom} W(x_{1},\ldots, x_{M}) = {1\over
2i\pi}\oint_{C_{0}} z^{-M-1}
E\Bigl(z,\sum_{i=1}^{M}x_{i}z^{i-1}\Bigr)\, \d z.\end{equation}
For example, in the case of one variable we obtain $W(x)=
E_{1}(x)$, and in the case of two variables
\be\label{genW2} W(x,y) = \sum_{i\geq 0}
E_{2-i}^{(i)}(x) y^{i}/i!\, .\end{equation}
It might seem that different geometries (\ref{geogen}) can yield the 
same $W$, but this is not the case. For example, in the 
case $M=2$, the holomorphic change of variables
\be\label{shift}\begin{split}
w_{2}&\rightarrow w_{2}-\sum_{i<0}E'_{2-i}(w_{1}) 
z^{-i-1}\, ,\\ w'_{2}&\rightarrow w'_{2} + \sum_{i\geq 2}
\sum_{j=0}^{i-2} E_{2-i}^{(j+1)}(0){w'}^{j}_{1}{z'}^{i-2-j}/j!\, ,
\end{split}\end{equation}
put the geometry in the form
\be\label{2matgeo} z'=1/z\, ,\quad w'_{1}=z^{-1}w_{1}\, , \quad
w'_{2}=z^{3}w_{2} + \sum_{i\geq 0}E'_{2-i}(w_{1})z^{2-i}\, ,\end{equation}
with the conditions
\be\label{consV} E_{2-i}^{(j)}(0) = 0 
\quad {\rm for}\quad 0\leq j\leq i-1\end{equation}
on the derivatives of the functions $E_{k}$. There is thus a unique 
geometry for a given superpotential.

\subsubsection{The non-commutative structure}
Equation (\ref{genW2}) shows that we can engineer an arbitrary
superpotential of two variables when only one brane is wrapped on
$\pone$. However, ordering ambiguities arise when $N\geq 2$ D5 branes
are considered. Eventually we expect that only special, in some sense
integrable, matrix models can be constructed. The rigorous computation
of the matrix model potential $W$ as a function of matrices $X_{i}$
would require an extensive use of the theory of branes in Calabi-Yau
threefolds, and this is beyong the scope of the present paper. There
is however little doubt as to what the correct answer must be. Our
proposal is that the parameters $x_{i}$ parametrizing the fluctuations
of the branes in (\ref{w1exp}) must be replaced by hermitian matrices
$X_{i}$. Supersymmetry then implies (\ref{cgen}) with
$x_{j}\rightarrow X_{j}$. Remarkably, with this prescription for the
ordering, the constraints are still integrable, and the potential is
given by (\ref{potential}).

In the two-matrix case, on which we shall mainly focus in the
following, the above prescription tells that the non-commutative
version of a monomial $x^{i}y^{j}$ is simply obtained by expanding
$\tr (X+Y)^{i+j}$ and picking up the terms with the right degrees in
$X$ and $Y$. For example, $x^{2}y^{2}\rightarrow\tr (2X^{2}Y^{2}/3 +
XYXY/3)$. The general formula is
\be\label{mcmono} x^{i}y^{j}\longrightarrow {i!j!\over (i+j)!} \oint_{C_{0}} 
{\d z\over 2i\pi}\, z^{-1-j} \tr (X + Y z)^{i+j}\, .\end{equation}
For polynomials $V_{X}$, $V_{Y}$ and $V$, the geometric potential
\be\label{Upot} E(z,w) = z^{2} V_{X}(w) + z^{2} V_{Y}(w/z) +
{z^{2}V(w)\over 1-1/z}\end{equation}
yields models of the form
\be\label{2mmmono} W(X,Y)=V_{X}(X) + V_{Y}(Y) + V(X+Y)\, .\end{equation}
The term $(1-1/z)^{-1}$ in (\ref{Upot}) must be understood as a
formal power series in $z^{-1}$, for which only the first $\deg V +1$
terms are relevant. Special cases of (\ref{2mmmono}) have been obtained
by considering Calabi-Yau threefolds that are monodromic fibrations of
ALE spaces with ADE singularities \cite{CKV}. The A-series correspond
to $\deg V=2$, the D-series to $\deg V=\deg V_{Y}=3$ and the E-series
to $\deg V=3$, $\deg V_{Y}=4$ and $4\leq\deg V_{X}\leq 6$. For those
models, the deformed geometry $\cal M$ can in principle be calculated by
using the results of \cite{KM}.

\subsection{Blowing down}
\subsubsection{Theorems}

Let us first state an important theorem by Laufer \cite{laufer}:

\noindent T{\scshape heorem} (Laufer):
Let ${\cal M}_{0}$ be an analytic space of dimension $D\geq 3$ with an
isolated singularity at $p$. Suppose that there exists a non-zero
holomorphic $D$-form $\Omega$ on ${\cal M}_{0}\backslash \{p\}$. Let
$\pi :\hat{\cal M}\rightarrow {\cal M}_{0}$ be a resolution of ${\cal
M}_{0}$. Suppose that the exceptional set $A=\pi^{-1}(p)$ is
one-dimensional and irreducible. Then $A$ is isomorphic to $\pone$ and
$D=3$. Moreover, the normal bundle of $\pone$ in
$\hat{\cal M}$ must be either ${\cal N}=\O(-1)\oplus\O(-1)$, or ${\cal
N}=\O(0)\oplus\O(-2)$, or ${\cal N}=\O(1)\oplus\O(-3)$.

\noindent It is known in general that $\cal N$ is a direct sum of
line bundles \cite{gro}, ${\cal N}=\O(n) \oplus\O(m)$. The condition
on the normal bundle in Theorem 1 is thus equivalent to asymptotic
freedom, $M\leq 2$. This is consistent with the fact that
asymptotically free theories are the one for which we can expect the
$\pone$ to be exceptional. The situation for $M\geq 3$ is more subtle.
An important point is that the normal bundle to the $\pone$ in the
geometry (\ref{geogen}) is changed when the perturbation $E$ is added.
From the gauge theory point of view, turning on the superpotential
amounts to giving a mass to the chiral multiplets, and generically the
theory flows to the pure gauge theory in the IR. More precisely, the 
number of massless chiral multiplets in a given vacuum is equal to 
the corank of the Hessian of $W$ at the corresponding critical point.
This translates mathematically in the following

\noindent C{\scshape onjecture} (RG flow): Consider the geometry
(\ref{geogen}) for $m=-n-2$ and associated superpotential $W$ given by
(\ref{potcom}). Let $\cal N$ be the normal bundle of a $\pone$ that
sits at a given critical point of $W$. Let $r$ be the corank of the
Hessian of $W$ at the critical point. Then ${\cal
N}=\O(r-1)\oplus\O(-r-1)$.

\noindent We give an elementary proof of this conjecture for $n=1$ in
the Appendix. This result suggests that models with an arbitrary
number of multiplets, or integrals over an arbitrary number of
matrices, may still be described by a geometric transition as in
Figure 3. The difficulty is that Laufer's theorem implies that the
blow down map must be singular when the parameters are adjusted in
such a way that the corank of the Hessian jumps to a value greater
than two. Because of this complication, we restrict ourselves to
two-matrix models in the following.

\subsubsection{Examples}

The blow down map $\pi:\hat{\cal M}\rightarrow {\cal M}_{0}$ is given
explicitly by four globally holomorphic functions $\pi_{i}$, $1\leq
i\leq 4$. The mapping $\pi$ must be a birational isomorphism except on
the $\pone$s that are mapped onto the singular points of ${\cal
M}_{0}$. We present the construction for three examples.

\noindent E{\scshape xample} 1: The resolved geometry
\be\label{res1mm} z'=1/z\, ,\ w'_{1} = w_{1}\, ,\
w'_{2}=z^{2}w_{2}+zP(w_{1})\, ,\end{equation}
engineers the one-matrix model with a potential $W(X)$ such that $W'=P$. 
The $\pone$s sit at
\be\label{p11mm} w_{1}(z)=x=w_{1}'(z')\, ,\quad w_{2}(z)=0=w_{2}'(z')\, 
,\end{equation}
with $P(x)=0$. The blow down is straightforwardly found to be
\ba &&\pi_{1}=w_{1}=w_{1}'\, ,\\
&&\pi_{2}=2zw_{2}+P(w_{1}) = 2z'w_{2}' - P(w_{1}')\, ,\\
&&\pi_{3}=w_{2}-zP(w_{1})-z^{2}w_{2}= -w_{2}' - 
z'P(w_{1}')+{z'}^{2}w_{2}'\, ,\label{p31mm}\\
&&\pi_{4}= w_{2}+zP(w_{1})+z^{2}w_{2}= w_{2}' - 
z'P(w_{1}')+{z'}^{2}w_{2}'\, .\label{p41mm}\ea
As required, $\pi=(\pi_{1},\pi_{2},\pi_{3},\pi_{4})$ maps the $\pone$s 
onto points,
\be\label{ponem1mm} \pi_{1}(\pone) = x\, ,\quad 
\pi_{2}(\pone)=\pi_{3}(\pone)=\pi_{4}(\pone)=0\, ,\end{equation}
and is a birational isomorphism outside the $\pone$s whose inverse is 
given by
\be\label{inv1mm} w_{1}=\pi_{1}\, ,\quad w_{2}={1\over 2} \left(\strut
\pi_{3}+\pi_{4}\right) ,\quad z = {\pi_{2}-P(\pi_{1})\over 
\pi_{3}+\pi_{4}}\,\cdotp\end{equation}
By taking the difference between (\ref{p31mm}) and (\ref{p41mm}),
we obtain the following algebraic constraint
\be\label{M01mm} {\cal M}_{0}:\ 
\pi_{4}^{2}=\pi_{3}^{2}+\pi_{2}^{2}-P^{2}(\pi_{1})\end{equation}
that defines the singular Calabi-Yau geometry.

\noindent E{\scshape xample} 2: The resolved geometry for the model 
(\ref{2mmpot}) takes the form
\be\label{res2mm} z'=1/z\, ,\ w'_{1} = z^{-1}w_{1}\, ,\
w'_{2}=z^{3}w_{2}+zQ(w_{1}/z) + z^{2}P(w_{1})-w_{1}z\, .\end{equation}
From (\ref{w1exp}) and (\ref{w2gen})
we see that the $\pone$s sit at
\ba\hskip -1.5cm && w_{1}(z)  = x + y z\, ,\quad  w'_{1}(z') = y + x 
z'\, ,\label{p2mm1}\\ \hskip -1.5cm &&
w_{2}(z)  = {P(x)-P(x+yz)\over z}\, \cvp\quad
w'_{2}(z')={Q(y+xz')-Q(y)\over z'}\,\cdotp\label{p2mm2}\ea
The blow down map can be constructed by trial and error. For example,
starting with the ansatz
\be\pi_{1}=-z'w'_{2} + \cdots = 
-z^{2}w_{2}-Q(w_{1}/z) + w_{1} - zP(w_{1}) + \cdots\, ,\end{equation}
we see that the missing part can be adjusted to cancel the
non-holomorphic piece $Q(w_{1}/z)$ in the right hand side, finally
yielding
\be\label{pi1}\pi_{1}=-z'w'_{2} + Q(w'_{1}) = w_{1}
-zP(w_{1})-z^{2}w_{2}\, .\end{equation}
One can construct similarly
\be\label{pi2}\pi_{2}= zw_{2}+P(w_{1})=w'_{1}-z'Q(w'_{1})+
{z'}^{2}w'_{2}\, .\end{equation}
The construction of $\pi_{3}$ and $\pi_{4}$ requires further thought. 
A hint is given by calculating $\pi_{1}$ and $\pi_{2}$ on the 
spheres (\ref{p2mm1}), (\ref{p2mm2}),
\be\label{pisp}\pi_{1}(\pone)=x\, ,\quad\pi_{2}(\pone)=y\, .\end{equation}
Consistently with the fact that the $\pone$s are mapped onto points,
the result is independent of $z$ or $z'$. The result also suggests that 
$P(\pi_{1})$ and $Q(\pi_{2})$ could be useful globally holomorphic 
objects to consider. One then finds that
\ba\hskip -1cm
\pi_{3}\!&=\!& -{z'}^{3}w'_{2}+{z'}^{2}Q(w'_{1}) + z'P(\pi_{1}) - z'w'_{1}
= {P(\pi_{1}) - P(w_{1})\over z} - w_{2}\, ,\label{pi3}\\ \hskip -1cm
\pi_{4}\!&=\!& z^{3}w_{2}+z^{2}P(w_{1}) + zQ(\pi_{2}) - 
zw_{1}= {Q(\pi_{2}) - Q(w'_{1})\over z'} +
w'_{2}\, ,\label{pi4}\ea
have all the required properties. In particular
\be\label{pisp2}\pi_{3}(\pone)=\pi_{4}(\pone)=0\end{equation}
are $z$-independent, and the inverse is given by
\be w_{1}=\pi_{1}+{\pi_{2}\pi_{4}\over Q(\pi_{2})-\pi_{1}}\, \cvp\quad
w_{2}={(\pi_{2} - P(w_{1}))\pi_{3}\over P(\pi_{1})-\pi_{2}}\, 
\cvp\quad z={\pi_{4}\over Q(\pi_{2})-\pi_{1}}\,\cvp\end{equation}
with similar formulas for the primed variables. The singular 
Calabi-Yau is then found to be
\be\label{s1CY}{\cal M}_{0}:\
\pi_{3}\pi_{4}=\left(\strut P(\pi_{1})-\pi_{2}\right)
\left(\strut Q(\pi_{2})-\pi_{1}\right) .\end{equation}

\noindent E{\scshape xample} 3: We consider the superpotential
\be\label{lll}W(X,Y) = XY^{2} + V(X) + YU(Y^{2})\, .\end{equation}
Note that a term in $Y^{2}$ could be generated by a simple shift in 
$X$. It is useful to introduce
\be\label{derdef}\begin{split}
V'(x) &= P(x) = -F_{1}(x^{2})- x F_{2}(x^{2})\, ,\\
Q(y) &= U(y^{2}) + 2y^{2}U'(y^{2}) = -G(y^{2})\, .\end{split}\end{equation}
The resolved geometry is
\be\label{reslll} z'=1/z\, ,\ w'_{1} = z^{-1}w_{1}\, ,\
w'_{2}=z^{3}w_{2}+zQ(w_{1}/z) + z^{2}P(w_{1})+w_{1}^{2}\, ,\end{equation}
and the equations for the spheres are
\ba\hskip -1.5cm && w_{1}(z)  = x + y z\, ,\quad  w'_{1}(z') = y + x 
z'\, ,\label{plll1}\\ \hskip -1.5cm &&
w_{2}(z)  = {P(x)-P(x+yz)\over z}\, \cvp\quad
w'_{2}(z')=x^{2}+{Q(y+xz')-Q(y)\over z'}\,\cdotp\label{plll2}\ea
The blow down map is constructed using the same tricks as for Example 1. 
A special case of the map also appears in \cite{laufer} and \cite{CKV}.
We find
\ba\hskip -.7cm
\pi_{1}\! &=&\! w_{1}^{2}-zG(\pi_{2})+z^{2}P(w_{1})+z^{3}w_{2}  =
w'_{2}+{G({w'}^{2}_{1})-G(\pi_{2})\over z'}\,\cvp\label{p1ll}\\
\hskip -.7cm\pi_{2}\! &=&\! -z w_{2} - P(w_{1}) 
={w'}^{2}_{1}+ z'Q(w'_{1})-{z'}^{2}w'_{2} \, ,
\label{p2ll}\\ \hskip -.7cm
\pi_{3}\! &=& \! z'\pi_{2} - w'_{1} F_{2}(\pi_{1}) - z' 
F_{1}(\pi_{1})\nonumber\\
\! &=&\!  -w_{2} + {F_{1}(w_{1}^{2}) - F_{1}(\pi_{1})+
w_{1}(F_{2}(w_{1}^{2}) - F_{2}(\pi_{1}))\over z}\,\cvp\label{p3ll}\\
\hskip -.7cm\pi_{4}\! &=& \!w'_{1}\pi_{2}-z'\pi_{1}F_{2}(\pi_{1})-w'_{1}F_{1}(\pi_{1})
\nonumber\\ \! &=& \!
-w_{1}w_{2} + {w_{1}(F_{1}(w_{1}^{2}) - F_{1}(\pi_{1}))+
w_{1}^{2} F_{2}(w_{1}^{2}) - \pi_{1}F_{2}(\pi_{1}))\over z}
\,\cdotp\label{p4ll}\ea
The blow down can be inverted using, for example, the relations
\ba z'\! &=& \! {(\pi_{2}-F_{1}(\pi_{1}))\pi_{3} + 
F_{2}(\pi_{1})\pi_{4}\over (\pi_{2} - 
F_{1}(\pi_{1}))^{2}-\pi_{1}F_{2}^{2}(\pi_{1})}\, \cvp\\
w'_{1}\! &=& \! 
{\pi_{1}\pi_{3}F_{2}(\pi_{1})+(\pi_{2}-F_{1}(\pi_{1}))\pi_{4}\over 
(\pi_{2} - F_{1}(\pi_{1}))^{2}-\pi_{1}F_{2}^{2}(\pi_{1})}\,\cvp\\
w'_{2}\! &=& \! \pi_{1}+ {G(\pi_{2})-G({w'}^{2}_{1})\over z'}\,\cdotp\ea
The $\pone$s are mapped onto points,
\be\label{spll}\pi_{1}(\pone) = x^{2}\, ,\ \pi_{2}(\pone) = 
y^{2}\, ,\ \pi_{3}(\pone)=-yF_{2}(x^{2})\, ,\ \pi_{4}(\pone) = 
xyF_{2}(x^{2}) ,\!\end{equation}
and the singular geometry is given by
\begin{multline} {\cal M}_{0}:\
\pi_{2}\Bigl[ \left(\strut \pi_{2}-F_{1}(\pi_{1})\right)^{2} - 
\pi_{1} F_{2}^{2}(\pi_{1})\Bigr] = \\ 
\pi_{4}^{2} - \pi_{1}\pi_{3}^{2} - 
G(\pi_{2})\Bigl[ \left(\strut \pi_{2}-F_{1}(\pi_{1})\right) \pi_{3} + 
\pi_{4}F_{2}(\pi_{1})\Bigr] .\label{s2CY}\end{multline}
\subsection{The singular geometry and privileged coordinates}

The singular points in a geometry
\be\label{sggeo}{\cal M}_{0}:\ 
{\cal G}_{0}(\pi_{1},\pi_{2},\pi_{3},\pi_{4})=0\end{equation}
are obtained by solving the equations
\be\label{sgeq} {\cal G}_{0}=0\, ,\quad \d {\cal G}_{0}=0\end{equation}
simultaneously. By construction, the points $\pi_{i}(\pone)$, which
are the images of the two-spheres, are singular. They are associated
with the one dimensional representations of the algebra of classical
equations of motion, since the $\pone$s sit at the extrema of the
superpotential $W$ as a function of commuting variables. A deep
consistency requirement is that there are also singular points
associated with the higher dimensional representations \cite{CKV},
because the geometry encodes the most general solution of the matrix
model. The full set of representations must thus be found from the
algebraic equations (\ref{sgeq}) over commuting variables. This is
possible if there is a correspondence between some combinations of the
coordinates and the Casimir operators that characterize the
representations. These combinations are related to the privileged
coordinates that we have already discussed in Section 2.3. There can
also be additional singularities, that are reminiscent of the commonly
found branch cuts on unphysical sheets for the resolvents.

\noindent E{\scshape xample} 1: The singular points of the geometry 
(\ref{M01mm}) are at
\be\label{sg1mm} P(\pi_{1})=0\, ,\quad \pi_{2}=\pi_{3}=\pi_{4}=0\, ,\end{equation}
which yields the obvious identification
\be\label{id1mm} X\equiv \pi_{1}=x\, .\end{equation}

\noindent E{\scshape xample} 2: For (\ref{s1CY}), equations 
(\ref{sgeq}) are equivalent to
\be\label{sl1} \pi_{1}=Q(\pi_{2})\, ,\quad \pi_{2}=P(\pi_{1})\, 
,\quad \pi_{3}=\pi_{4}=0\, .\end{equation}
This is perfectly consistent with (\ref{2mmirrep}), with the identification
\be\label{pri1} X\equiv \pi_{1}=x\, ,\quad Y\equiv\pi_{2}=y\, ,\end{equation}
that is also suggested by (\ref{pisp}). 

\noindent E{\scshape xample} 3: The algebra of classical equations of
motion associated with the model (\ref{lll}),
\be\label{algLgen}{\cal A}:\ \{ X,Y\} = G(Y^{2})\, ,\quad 
Y^{2}=F_{1}(X^{2})+XF_{2}(X^{2})\, ,\end{equation}
is very similar to the one studied in Section 2.2.3, and provides
an example with both one and two dimensional representations. The
generic singular points of the associated ${\cal M}_{0}$ (\ref{s2CY})
are found to satisfy either
\ba\label{sglll1} && 4\pi_{1}\pi_{2}=G^{2}(\pi_{2})\, ,\quad
\left(\strut
\pi_{2}-F_{1}(\pi_{1})\right)^{2}=\pi_{1}F_{2}^{2}(\pi_{1})\, 
,\nonumber\\
&& \pi_{3}={(F_{1}(\pi_{1})-\pi_{2})G(\pi_{2})\over 
2\pi_{1}}\,\cvp\quad \pi_{4}={1\over 2} G(\pi_{2})F_{2}(\pi_{1})\, ,\ea
which yield the one dimensional representations, or
\be\label{sglll2} F_{2}(\pi_{1})=0\, ,\quad 
\pi_{2}=F_{1}(\pi_{1})\, ,\quad \pi_{3}=\pi_{4}=0\, ,\end{equation}
which yield the two dimensional representations. We obtain the 
identifications
\be\label{pri2} X^{2}\equiv \pi_{1}=x^{2}\, ,\quad Y^{2}\equiv 
\pi_{2}=y^{2}\, ,\end{equation}
which also make (\ref{sglll1}) and (\ref{sglll2})
perfectly consistent with (\ref{spll}).

\subsection{The deformed geometry}

Singularities of ${\cal M}_{0}$ lie at the classical critical
points of the superpotential. When the coupling $S$ is turned on, the
eigenvalues spread and the critical points are replaced by cuts. The
defining equation (\ref{sggeo}) of ${\cal M}_{0}$ is then deformed in
such a way that the singularities are replaced by three-spheres. The 
equation of the deformed space is of the form
\begin{multline}
\label{defsp}{\cal M}:\ {\cal G}(\pi_{1},\pi_{2},\pi_{3},\pi_{4}) = \\
{\cal G}_{0}(\pi_{1},\pi_{2},\pi_{3},\pi_{4}) + S\!\!\!\!\!
\sum_{(a,b,c,d)\in{\mathbb 
N}^{4}}c_{abcd}\, \pi_{1}^{a}\pi_{2}^{b}\pi_{3}^{c}\pi_{4}^{d}=0\, .
\end{multline}
To find the allowed monomials
$\pi_{1}^{a}\pi_{2}^{b}\pi_{3}^{c}\pi_{4}^{d}$, we must impose that
the only divergences that occur in the period integrals (\ref{solCY1})
are either $S$-independent or linear in $S$ and logarithmic. This
constraint comes from the renormalizability of the gauge theory on the
brane. A logarithmic divergence proportional to $S$ is absorbed by the
standard renormalization of the gauge coupling constant. It is
important to realize that divergences proportional to higher powers of
$S$ are not allowed, even if they are logarithmic. For example, this
subtlety must be taken into account to find the correct deformation of
(\ref{s2CY}). Another constraint, from analyticity, is that a monomial
that is generically forbidden cannot appear in a special case, for
example when a coupling in the matrix model potential is set to zero.
After all those constraints have been taken into account, and up to
coordinate redefinitions, the number of deformation parameters
$c_{abcd}$ must match the number of irreducible representations,
because there should be a uniquely defined deformed space for each 
vacuum (\ref{vac}).

It is usually straightforward to guess the 
general form of the deformations up to coordinate transformations. 
For example, the deformed version of (\ref{M01mm}),
(\ref{s1CY}) and (\ref{s2CY}) are respectively
\ba&&\hskip -1.6cm {\cal M}:\ 
\pi_{4}^{2}=\pi_{3}^{2}+\pi_{2}^{2}-P^{2}(\pi_{1}) + 
S\Delta(\pi_{1})\, ,\label{M1mm}\\
&&\hskip -1.6cm{\cal M}:\
\pi_{3}\pi_{4}=\left(\strut P(\pi_{1})-\pi_{2}\right)
\left(\strut Q(\pi_{2})-\pi_{1}\right) + 
S\Delta(\pi_{1},\pi_{2})\, ,\label{M1}\\
&& \hskip -1.6cm{\cal M}:\
\pi_{2}\Bigl[ \left(\strut \pi_{2}-F_{1}(\pi_{1})\right)^{2} - 
\pi_{1} F_{2}^{2}(\pi_{1})\Bigr]+S\Delta(\pi_{1},\pi_{2})
= \nonumber\\ &&\hskip -1.3cm
\pi_{4}^{2} - G(\pi_{2})F_{2}(\pi_{1})\pi_{4}
- \pi_{1}\pi_{3}^{2} - 
\Bigl[ G(\pi_{2}) \left(\strut \pi_{2}-F_{1}(\pi_{1})\right) + 
S\hat\Delta(\pi_{2})\Bigr]\pi_{3}\, ,\label{M2}\ea
where $\Delta$ and $\hat\Delta$ are polynomials. The constraints on
the degrees of the polynomials can be found by explicit calculations.
For the one-matrix model (\ref{M1mm}) it is well-known that
$\deg\Delta = \deg P -1$. For the model (\ref{M1}), we will find in
the next Section that the most general monomial in $\Delta$ is
$\pi_{1}^{a}\pi_{2}^{b}$ with $0\leq a\leq\deg P-1=d_{X}-1$ and $0\leq
b\leq\deg Q-1=d_{Y}-1$. The number of deformation parameters is thus
$d_{X}d_{Y}$, consistently with (\ref{2mmirrep}). In the case of
(\ref{M2}), three cases must be distinguished. If $d_{G}=\deg G=0$,
$\hat\Delta$ must be a constant and the allowed monomials in $\Delta$
are $\pi_{1}^{a}$ for $0\leq a\leq d-1$ and $\pi_{1}^{a}\pi_{2}$ for
$0\leq a\leq [(d-1)/2]$. If $d_{G}=d=1$, $\hat\Delta$ is a
constant and $\Delta$ is linear in $\pi_{2}$ and independent of
$\pi_{1}$. In all the other cases, $\deg\hat\Delta = d_{G}-1$ and
$\Delta$ is a linear combination of terms $\pi_{1}^{a}\pi_{2}^{b}$
with $0\leq a\leq d-1$ for $0\leq b\leq d_{G}-1$, $0\leq a\leq d-2$
for $d_{G}\leq b\leq 2d_{G}-1$, and $0\leq a\leq [(d-1)/2]-1$ for
$b=2d_{G}$. It is easily to check that the total number of
parameters is given by $\max (d+2,2dd_{G}) + [(d-1)/2]$,
matching the number of irreducible representations of the algebra
(\ref{algLgen}).

\section{Solutions from the geometry}
\setcounter{equation}{0}

The aim of this Section is twofold. First we want to explain how to get
geometrically the resolvents for various matrices. Second we provide
full calculations for the models (\ref{2mmpot}) and (\ref{lll}). The
solution of this latter example was not known, but we will be able to
check the results in a particular case in the next Section. Let us
emphasize that any model of the form (\ref{potential}), at least for
two matrices, should be in principle solvable using the same strategy.

\subsection{The resolvents from the geometry}

The idea is that, since any multi-matrix model can be formulated as a
one-matrix model by integrating out all but one matrix, the Calabi-Yau
geometry for a general model should have some common features with the
geometry (\ref{M1mm}) for the one-matrix model. The relevant property 
of the latter geometry is that it is a fibration of
the deformation of the simplest ALE space
${\mathbb C}^{2}/{\mathbb Z}_{2}$,
\be\label{ALE} u^{2}=v^{2}+w^{2}+\lambda (z)\, .\end{equation}
The space $u^{2}=v^{2}+w^{2}+\lambda$ has a single $S^{2}$ of
holomorphic volume $\lambda$. When $\lambda(z)=0$, the two-sphere
shrinks. Classically this is equivalent to the equations of motion.
The base coordinate $z$ describes the fluctuations of $S^{2}$ and is
associated with the matrix $X$, $z=x$ as in (\ref{id1mm}). We thus
expect that the more general spaces we have to deal with are natural
fibrations over bases parametrized by the privileged coordinates
discussed in Section 3.3. The fiber $F_{x}$ over a point $x$ is in
general much more complicated than a simple deformed ${\mathbb
C}^{2}/{\mathbb Z}_{2}$ space. For example, the multi-valuedness of
the effective potential implies that the fibers must contain several
$S^{2}$s.

\begin{figure}
\centerline{\epsfig{file=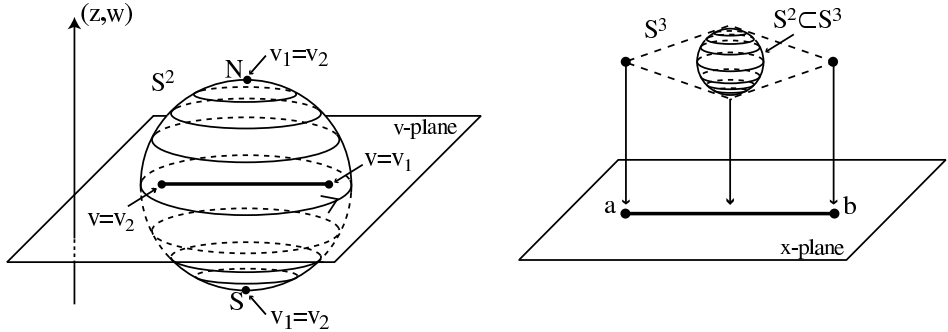,width=12.5cm}}
\caption{Integrating over $S^{3}$. We first integrate over
$S^{2}\subset S^{3}$ (left inset) and then over a meridian of $S^{3}$
(right inset). From the matrix model perspective, the meridian $[a,b]$
is either an eigenvalue-filled interval corresponding to a branch cut in
Figure 1, or to a intervals of the type $[b_{K},a_{K'}]$ or 
$]\mu_{0},b_{K}]$.
\label{fig4}}
\end{figure}

To make those ideas quantitative, we propose that integrating over the
two-spheres in the fibers yields the discontinuity of the resolvent
across a branch cut. Let us introduce the coordinate $x$ associated
with the matrix $X$, and let us denote by ${\rm i}_{X}$ the interior
product associated with the vector field $\partial/\partial x$. If
$g^{X}$ is the resolvent for $X$ on the physical sheet, and $\hat
g^{X}$ is the analytic continuation of $g^{X}$ through a branch cut,
then
\be\label{S2i} \int_{S^{2}\subset F_{x}}{\rm i}_{X}\Omega = 
g^{X}(x)-\hat g^{X}(x)\, .\end{equation}
This is the fundamental equation that unambiguously determines
$g^{X}$, as can be seen for example by taking $x$ to be in the support
of $\rho^{X}$ and using (\ref{densfor}) and (\ref{resfor}).

We can now integrate over a three-sphere in the full three-fold
geometry by first integrating over the $S^{2}$ in the fiber and then
over a meridian $[a,b]$ of $S^{3}$ in the base, as depicted in Figure
4. Equation (\ref{S2i}) implies that
\be\label{sgOmega}\oint_{S^{3}}\Omega = 
\int_{a}^{b}\int_{S^{2}\subset F_{x}}\Omega=\int_{a}^{b}
\left( g^{X}(x)-\hat g^{X}(x)\right)\d x = \oint g^{X}(x)\,\d x\, ,\end{equation}
which links the special geometry relations in the Calabi-Yau
(\ref{solCY1}) and matrix model (\ref{sgea}), (\ref{sgeb}) forms.

As we have said, there are in general several $S^{2}\subset F_{x}$
over which to integrate. The relevant $S^{2}$s can be easily identified
by looking at the classical $S\rightarrow 0$ limit of (\ref{S2i}).
Equations (\ref{S2i}) and (\ref{gsp}) imply
\be\label{cleff} 
\lim_{S\rightarrow 0}\int_{S^{2}\subset F_{x}}{\rm 
i}_{X}\Omega = f_{\rm b,\, cl}(x)\, ,\end{equation}
where the background force was defined in (\ref{forcegdef}) (for the
$D\geq 2$ dimensional representations, one must take into account the
terms discussed in Section 2.2.3 in the classical background force).
Note that a rigorous consistency condition that follows from the second
equation in (\ref{solCY1}) is that the non-compact integrals of
$\lim_{S\rightarrow 0}\int_{S^{2}\subset F_{x}}{\rm i}_{X}\Omega$ are
related to the critical values of the potential $W$. We believe that
the stronger condition (\ref{cleff}) holds if and only if the ansatz
(\ref{S2i}) is correct. Equation (\ref{cleff}) is very handy to fix the
overall normalization of $\Omega$. Let us also point out that
the classical equations of motion, including the higher dimensional
representations, are strictly equivalent to the condition $f_{\rm b,\,
cl}(x)=0$,
\be\label{clS21} \tr\d W = 0\Longleftrightarrow
\lim_{S\rightarrow 0}\int_{S^{2}\subset F_{x}}{\rm 
i}_{X}\Omega = 0\, .\end{equation}
The basic requirement of renormalizability is equivalent to the fact
that the quantum corrections to $\int_{S^{2}\subset F_{x}}{\rm
i}_{X}\Omega$ must vanish faster than $1/x$ when $x\rightarrow\infty$,
except for terms linear in $S$ that may go as $1/x$. As explained in
Section 3.4, this must restrict the number of moduli of the deformed
space to be equal to the number of irreducible representations of
${\cal A}$.

\noindent E{\scshape xample}: 
Let us consider a non-compact Calabi-Yau threefold in ${\mathbb C}^{4}$
defined by the equation
\be\label{genCY} {\cal M}:\ u^{2} = f(z,w)v^{2}+g(z,w)v+h(z,w)\, ,\end{equation}
where $f$, $g$ and $h$ are polynomials. 
The nowhere vanishing holomorphic three-form is
\be\label{genOmega} \Omega = N\, {\d v\wedge\d z\wedge\d w\over 2i\pi
u}\,\cvp\end{equation}
where $N$ is a normalization constant. We assume that the privileged
coordinate $x$ is expressed in terms of the variables $z$ and $w$
only, so that
\be\label{iXO} {\rm i}_{X}\Omega = -N\, {\d v\over 2i\pi u}\wedge
{\rm i}_{X}(\d z\wedge\d w)\, .\end{equation}
As in Figure 4, 
the two-spheres are generated by closed contours encircling the branch 
cut $[v_{1},v_{2}]$ in the $v$-plane. Equation (\ref{S2i}) then yields
\ba g^{X}(x)-\hat g^{X}(x) &\!=\!& -{N\over 2i\pi}\int_{\rm 
S}^{\rm N} \oint {\d v\over\sqrt{(v-v_{1})(v-v_{2})}}\wedge
{{\rm i}_{X}(\d z\wedge\d w)\over\sqrt{f(z,w)}}\nonumber\\ &\!=\!&
-N\int_{\rm S}^{\rm N}{{\rm i}_{X}(\d z\wedge\d w)
\over\sqrt{f(z,w)}}\,\cdotp\label{S2int}\ea
The North and South poles N and S lie on the algebraic curve 
$v_{1}=v_{2}$,
\be\label{AC}{\cal C}:\ g^{2}(z,w) - 4 f(z,w)h(z,w) = 0\, .\end{equation}
\subsection{The two-matrix model with $XY$ interaction}

By introducing the privileged coordinates (\ref{pri1}) $z=\pi_{1}=x$
and $w=\pi_{2}=y$ associated with the matrices $X$ and $Y$
respectively, the geometry (\ref{M1}) can be put in the form
\be\label{M1c} {\cal M}:\ u^{2}=v^{2} + {\cal C}(x,y)=v^{2}
+\left(\strut P(x)-y\right)\left(\strut Q(y)-x\right) + 
S\Delta(x,y)\, .\end{equation}
Equation (\ref{S2int}) yields
\be\label{2mmsol1} g^{X}(x) - \hat g^{X}(x) =-N \int_{\rm S}^{\rm N}\d 
y\, .\end{equation}
Let us denote by $y=y_{i}(x)$ the $d_{Y}+1$ solutions to the equation
\be\label{CAcons} {\cal C}\left(\strut x,y(x)\right) = 0\, ,\end{equation}
with labels chosen in such a way that, classically,
$y_{1,\,\rm cl}(x) = P(x)$ and $y_{i\, ,\rm cl}(x) =
Q^{-1}(x)$ for $i\geq 2$. Consistency with (\ref{cleff}) and 
(\ref{fcl2mm}) then requires that $N=\pm 1$ and
\be\label{2mmsol2}  g^{X}(x) - \hat g^{X}(x) = y_{i}(x) - y_{1}(x)\, 
.\end{equation}
Let us note that the physical sheet branch cuts of the resolvent
correspond to and only to $y_{1}=y_{i}$, $i\geq 2$. From
(\ref{2mmsol2}) we thus deduce that $g^{X}$ and $-y_{1}$ have exactly
the same branch cuts and discontinuity across branch cuts, and thus
must be equal up to some entire function. From (\ref{asy}) we finally get
\be\label{2mmsol3} g^{X}(x) = P(x) -y_{1}(x)\, .\end{equation}
The same analysis could be repeated to compute the resolvent 
for the matrix $Y$.

Let us now implement the constraint of renormalizability, which
restricts the large $x$ behaviour of $y_{i}(x)-y_{1}(x)$. These
restrictions actually apply to the $S$-dependent part of $y_{1}(x)$
and $y_{i}(x)$ separately, because generically no cancellation can
occur. A straightforward calculation then shows that the allowed
monomials $x^{a}y^{b}$ in $\Delta(x,y)$ are such that
\be\label{abc1}
a+bd_{X}\leq d_{X}d_{Y}-1\, ,\quad a+b/d^{Y}\leq
d_{X}+(d_{Y}-1)/d_{Y}-1\, .\end{equation}
This is equivalent to $a\leq d_{X}-1$ and $b\leq d_{Y}-1$, showing that
there are exactly $d_{X}d_{Y}$ free parameters in $\Delta$,
matching the number of irreducible representations of $\cal 
A$.\goodbreak

\noindent S{\scshape ummary}: If ${\cal C}(x,y)=
\left(\strut P(x)-y\right)\left(\strut Q(y)-x\right) + 
S\Delta(x,y)$, with
\be\label{Delfor}\Delta(x,y) = 
\sum_{i=0}^{d_{X}-1}\sum_{j=0}^{d_{Y}-1}c_{ij}x^{i}y^{j}\, ,\end{equation}
then the resolvents $g^{X}$ and $g^{Y}$ for the model (\ref{2mmpot})
satisfy
\be\label{SOL2mm} {\cal C}\left(x,P(x)-g^{X}(x)\right) = {\cal C}
\left(Q(y)-g^{Y}(y),y\right) = 0\, ,\end{equation}
with the asymptotics (\ref{asy}). The $d_{X}d_{Y}$ parameters 
$c_{ij}$ are determined by a choice of vacuum through the $d_{X}d_{Y}$
conditions (\ref{sgea}).

The above result is in perfect agreement with the known 
solution from the loop equations \cite{2mm}, which yields
\be\label{Deltaloop1}\Delta(x,y) = 1 - \Bigl\langle {\tr\over n}\, 
{P(x)-P(X)\over x-X}{Q(y)-Q(Y)\over y-Y} \Bigr\rangle\, .\end{equation}
Note that in the loop equation approach the vacua are naturally
characterized by a set of basic correlators, whereas in the geometric
approach the natural parameters are the filling fractions
(\ref{Skdef}). There is a one-to-one correspondence between those two 
sets of parameters.

\subsection{The generalized Laufer's matrix model}

We now solve the model (\ref{lll}), with $\deg V'=d$ and $\deg
U=d_{G}$. When $U=0$ and $V(x)=x^{q}$, the relevant Calabi-Yau
(\ref{M2}) reduces to the geometry originally constructed by Laufer in
\cite{laufer}. With an obvious redefinition of coordinates, (\ref{M2})
can be cast in the form (\ref{genCY}),
\begin{multline} {\cal M}:\ u^{2} = zv^{2}+ \Bigl[
\left(\strut w - F_{1}(z)\right)G(w) + S\hat\Delta(w)\Bigr]
v\\  + {1\over 4}G^{2}(w)F_{2}^{2}(z) + w\Bigl[
\left(\strut w-F_{1}(z)\right)^{2}-zF_{2}^{2}(z)\Bigr] +
S\Delta(z,w)\, ,\label{M2c}\end{multline}
and the algebraic curve (\ref{AC}) is given by the equation
\begin{multline} {\cal C}(z,w) = \Bigl( zw-{G^{2}(w)\over
4}\Bigr)\Bigl( \left(\strut
w-F_{1}(z)\right)^{2}-zF_{2}^{2}(z)\Bigr)\\ +
Sz\Delta(z,w) - {1\over 2} S \left(\strut
w-F_{1}(z)\right)G(w)\hat\Delta(w) - {1\over 4} S^{2}\hat\Delta^{2}(w) =
0\, .\label{ACl}\end{multline}
\subsubsection{The resolvent for $X$}
By using the privileged coordinates (\ref{pri2}) $z=\pi_{1}=x^{2}$ and 
(\ref{S2int}) we get
\be\label{gXL1} g^{X}(x) -\hat g^{X}(x) = -2N \Delta w(x)\, ,\end{equation}
where $\Delta w(x)$ denotes the difference between two roots of the
equation
\be\label{CL1}{\cal C}\left( \strut x^{2},w(x)\right) =0\, .\end{equation}
There are $\max (3,2+2d_{G})$ roots, labeled such that classically
\begin{multline}\label{clL1}
w_{1,\, {\rm cl}}(x) = -V'(x)\, ,\quad w_{2,\, {\rm 
cl}}(x) = -V'(-x)\, ,\\ 4x^{2}w_{i,\, \rm cl}(x) = G^{2}\left(\strut 
w_{i,\,\rm cl}(x)\right)\quad {\rm for}\ i\geq 3\, .\end{multline}
The classical background force for one dimensional representations is
simply $f_{\rm b,\, cl}(x)=-y^{2}(x)-V'(x)$ with
$4x^{2}y^{2}(x)=G^{2}(y^{2}(x))$. Consistency with (\ref{cleff}) and
(\ref{clS21}) then immediately requires that $N=\pm 1/2$ and
\be\label{Lsol1} g^{X}(x)-\hat g^{X}(x) = w_{1}(x)-w_{i}(x)\, ,\quad
i\geq 2\, ,\end{equation}
where the cases $i=2$ and $i\geq 3$ correspond to two and one dimensional 
representations respectively. Finally the branch cut structure and 
asymptotics of the resolvent imply
\be\label{LsolX} g^{X}(x) = w_{1}(x) + V'(x)\, .\end{equation}

Let us now study the normalizability constraints that follow from
(\ref{Lsol1}). One must be careful that the term in $S^{2}$ in
(\ref{ACl}) does not produce any divergences, while terms in $S$ may
yield logarithmic divergences. The relevant constraints on the
monomials $z^{a}w^{b}$ in $\Delta (z,w)$ read
\ba \hskip -1cm 2 a + bd &\! \leq\! & \max(d+2,2dd_{G}) + 2 [(d-1)/2] 
-2\, ,\label{cLD1}\\ \hskip -1cm
(2d_{G} -1)a + b &\! \leq\! & (2d_{G} -1)d - d_{G}\ {\rm for}\ 
d_{G}\geq 1\ {\rm and}\ d\geq 2\ {\rm if}\ d_{G}=1\, ,\label{cLD2}\\
\hskip -1cm
a+b &\! \leq\! & 1\ {\rm for}\ d=d_{G}=1\, ,\label{cLD3}\ea
and the relevant constraints on $\hat\Delta(w)$ read
\ba 2 d\deg\hat\Delta &\! \leq\! & \max (d+2,2dd_{G}) + 2 [(d-1)/2] 
-1\, ,\label{cLD4}\\
d\deg\hat\Delta &\! \leq\! & \max (d+2,2dd_{G}) + 2 [(d-1)/2] - 
d(1+d_{G})\, .\label{cLD5}\ea
The inequalities (\ref{cLD1}), (\ref{cLD4}) and (\ref{cLD5}) are
obtained by studying the asymptotics of the roots $w_{1}$ or $w_{2}$,
and (\ref{cLD2}) and (\ref{cLD3}) follow from looking at $w_{i}$ for
$i\geq 3$. It is straightforward to check that the general solution to
this set of inequalities yields exactly the deformed geometry
described at the end of Section 3.4.

\noindent S{\scshape ummary}: Let $\cal C$ be defined by (\ref{ACl}),
with the constraints on $\Delta$ and $\hat\Delta$ as in Section 3.4.
The resolvent $g^{X}$ for the model (\ref{lll}) satisfies the degree
$\max(3,2+2d_{G})$ algebraic equation
\be\label{SOLLX} {\cal C}\left(x^{2},g^{X}(x)-V'(x)\right) = 0\, ,\end{equation}
with the asymptotics (\ref{asy}).

\subsubsection{The resolvent for $Y$}

Using (\ref{pri2}), (\ref{S2int}), and the normalization constant $N$ 
deduced in the preceding subsection, we obtain
\be\label{solLY} g^{Y}(y)-\hat g^{Y}(y) = 2 y\, \Delta 
x(y)\, ,\end{equation}
where $\Delta x(y)$
denotes the difference between two roots of the equation
\be\label{CL2} {\cal C}\left(\strut x(y)^{2},y^{2}\right)
= 0\, .\end{equation}
Let us separate the $2(d+1)$ roots into two sets $\{x_{i}\}$
and $\{-x_{i}\}$ characterized by
\be\label{cliLY} x_{1,\,\rm cl}(y) = -{G(y^{2})\over 2y}\,\cvp\quad
V'\left(\strut x_{i,\,\rm cl}(y)\right) + y^{2} = 0\ {\rm for}\ i\geq 
2\, .\end{equation}
One and two dimensional representations correspond respectively to
$x_{1}=-x_{i}$ and $x_{j}=-x_{i}$ for $i,j\geq 2$. From the classical
background force in one dimensional representations, $f_{\rm b,\,
cl}=G(y^{2})-2y x(y)$ with $y^{2}+V'(x(y))=0$, we deduce that
\be\label{difgY} g^{Y}(y) - \hat g^{Y}(y) = 
-2y\left(\strut x_{i}(y) + x_{j}(y)\right)\, ,\end{equation}
where $i\geq 2$ and $j\geq 1$. The resolvent $g^{Y}$ is the only
analytic function satisfying (\ref{difgY}), with no other branch cuts
than those associated with the representations of $\cal A$, and that
goes as $S/y$ at large $y$ on the physical sheet. A na\"\i ve guess
might have been to identify $g^{Y}$ with one of the root $x_{i}$ for
$i\geq 2$, but this cannot work since for example there would be
unphysical branch cuts corresponding to the permutation of the root
$x_{i}$ with another root $x_{j}$. It is easy to eliminate
those latter branch cuts by taking a permutation invariant
sum, normalized in such a way that (\ref{difgY}) is satisfied,
\be\label{gYL1} g^{Y}(y) = -y\sum_{i=1}^{d+1}x_{i}(y) + {\rm 
polynomial}\, .\end{equation}
The only possible remaining unwanted branch cuts in $\sum_{i} x_{i}$
would correspond to the permutation of $x_{1}$ with $-x_{1}$, but it
is actually straighforward to show that there are no cuts permuting
any root $x_{i}$ with its opposite. Indeed, the associated $S=0$
double points do not open up because the curve (\ref{ACl}) factorizes
at $z=0$ for any $S$,
\be\label{Clfact} {\cal C}(0,w) = -{1\over 4}\Bigl[
\left(\strut w-F_{1}(0)\right) G(w) + S\hat\Delta 
(w) \Bigr]^{2}\, .\end{equation}
Finally, we fix the polynomial part in (\ref{gYL1}) by looking at
the large $y$ asymptotics.

\noindent S{\scshape ummary}: Suppose that $V'(x) = \sum_{k=0}^{d} 
t_{k+1}x^{k}$. The resolvent
$g^{Y}$ for the model (\ref{lll}) is given by
\ba g^{Y}(y) &\! = \! & - y\sum_{i=1}^{d+1} x_{i}(y) -{1\over 2}
G(y^{2}) - {y t_{d}\over t_{d+1}}\quad {\rm if}\ d\geq
2\, ,\label{SOLLY1}\\ &\! = \! & - y \left(\strut x_{1}(y) +
x_{2}(y)\right) - {1\over 2} G(y^{2}) - {y (t_{1}+y^{2})\over t_{2}}
\quad {\rm if}\ d=1\, ,\label{SOLLY2}\ea
where the sum is taken over half of the roots of the equation 
(\ref{CL2}) satisfying the conditions (\ref{cliLY}).

The resolvent $g^{Y}$, being a sum of algebraic functions, must itself
be an algebraic function. It is not difficult to find the algebraic
equation satisfied by $g^{Y}$. Let us introduce the symmetric
polynomials $\sigma_{0}=1$ and $\sigma_{k}=\sum x_{i_{1}}\cdots
x_{i_{k}}$ for $1\leq k\leq d+1$. The resolvent is essentially
$-y\sigma_{1}$. We can write
\ba {\cal C}(x^{2},y^{2}) &\! = \!& (-1)^{d}t_{d+1}^{2}y^{2}
\prod_{i=1}^{d+1}\left(\strut x-x_{i}(y)\right)\left(\strut 
x+x_{i}(y)\right)\nonumber\\
&\! = \!& (-1)^{d}t_{d+1}^{2}y^{2} \sum_{0\leq k,k'\leq d+1} 
(-1)^{k}\sigma_{k}\sigma_{k'} x^{2d+2-k-k'}\, .\label{expC}\ea
Comparing with (\ref{ACl}), we get $d+1$ quadratic equations for the
$d+1$ unknown $\sigma_{k}$. By looking at the constant term and the
classical limit (\ref{cliLY}), we find that $\sigma_{d+1} =
(-1)^{d+1}\left(\strut
G(y^{2})(t_{1}+y^{2})+S\smash{\hat\Delta}(y^{2}) \right) /(2 y
t_{d+1})$. There remains $d$ quadratic equations for
$\sigma_{1},\ldots ,\sigma_{d}$. Geometrically, this means that the
resolvent lies at the intersection of $d$ quadrics. By elimination of
variables, we can find a degree $2^{d}$ algebraic equation satisfied
by $g^{Y}$. For example, when $d=1$, we recover the hyperelliptic
curve of the ordinary one-matrix model obtained by integrating out
$X$.

\section{Solution from the loop equations}
\setcounter{equation}{0}

The general class of models (\ref{potential}) ought to have some very
special mathematical properties that make possible a solution in terms
of an algebraic variety. In some sense, the loop equations should be
integrable in the planar limit. A general analysis is far beyong the
scope of the present paper, but we have been able to solve the example
(\ref{ldef}), which is a particular case of (\ref{lll}) for
$G=-\alpha$. We find a precise match with the result of the previous
Section, providing a very non-trivial test of the geometric approach,
and, most importantly, of the underlying conjectures on which it is
based. The test is particularly stringent because the model
(\ref{ldef}) is associated with a genuine ${\cal N}=1$ gauge theory
that is not a deformation of an underlying ${\cal N}=2$ theory. In
particular, the solution cannot be found by perturbing a
Seiberg-Witten curve, unlike all the examples studied so far in the
literature.

\subsection{General loop equations}

We consider the matrix integral
\be\label{Lmatint} \int\!\d X\d Y e^{-{n\over S} \tr \left(
XY^{2}+\alpha Y + V(X)\right)}\end{equation}
for an arbitrary polynomial
\be\label{Vesp} V(X) = \sum_{k=0}^{d+1}t_{k}X^{k}/k\, .\end{equation}
The basic object we are going to compute is the generating function
\be\label{defG}\Gamma(x,y) = S\, \Bigl\langle {\tr\over n}\, {1\over 
x-X}{1\over y-Y}\Bigr\rangle\, .\end{equation}
It can be expanded, either at large $y$ or at large $x$,
\be\label{Gexp} \Gamma (x,y) = \sum_{k\geq 0}g^{X}_{k}(x) y^{-k-1}
= \sum_{k\geq 0}g^{Y}_{k}(y) x^{-k-1}\, ,\end{equation}
in terms of generalized resolvents
\be\label{gXYkdef} g^{X}_{k}(x) = S\, \Bigl\langle {\tr\over n}\,
{Y^{k}\over x-X}\Bigr\rangle\,\cvp\quad g^{Y}_{k}(y) = S\,
\Bigl\langle {\tr\over n}\, {X^{k}\over y-Y}\Bigr\rangle\, .\end{equation}
The ordinary resolvents are $g^{X}=g^{X}_{0}$ and $g^{Y}=g^{Y}_{0}$.

Let us now consider the following variations in (\ref{Lmatint}),
\ba\hskip -1.5cm && \delta X = 0\, ,\quad \delta Y = {\epsilon\over 
x-X}\,\cvp\label{cvar0}\\ \hskip -1.5cm
&& \delta X = {\epsilon\over 2}
\Bigl( {1\over x-X} {1\over y-Y} + {1\over y-Y}{1\over x-X} 
\Bigr)\, ,\quad \delta Y=0\, , \label{cvar1}\\ \hskip -1.5cm
&& \delta X=0\, ,\quad \delta Y ={\epsilon\over 2}
\Bigl( {1\over x-X}{1\over y-Y}{1\over -x-X} + {1\over -x-X}{1\over 
y-Y}{1\over x-X} \Bigr)\, .\label{cvar2}\ea
The remarkable property of these changes of variables is that, due to
many cancellations of terms, the associated loop equations close in
the planar limit under the correlators of the form $\langle\tr
X^{p}Y^{q}\rangle$. After calculating the jacobian of the
transformations, taking into account the variation of the classical
potential, and using the factorization of multi-trace correlators at
large $N$, we get
\begin{multline}
\left(\strut g^{X}(x) - y^{2}-V'(x)\right) \Gamma (x,y) =
-\Bigl( y - {\alpha\over 2x}\Bigr) g^{X}(x)
- S\, {\langle \tr Y\rangle\over n x} \\
- S\, \Bigl\langle {\tr\over n}\, 
{1\over y-Y}{V'(x)-V'(X)\over x-X}\Bigr\rangle\label{le1}\,\cvp
\end{multline}
\begin{multline}
\Gamma (x,y)\Gamma (-x,y) = 
g^{X}(x) + g^{X}(-x) - y\left(\strut
\Gamma (x,y) + \Gamma (-x,y)\right)\\
- {\alpha\over 2x}\left(\strut
\Gamma (x,y) - \Gamma (-x,y)\right)\, .\label{le2}\end{multline}
\subsection{The resolvent for $X$}

The functions $g^{X}_{k}$, and in particular the resolvent for $X$, 
are obtained by expanding the loop equations at large $y$. 
From (\ref{le1}) we get the following recursion relations,
\ba g^{X}_{1}(x) &\!=\!& S\, {\langle \tr Y\rangle\over n x} - 
{\alpha\over 2x}\, g^{X}(x)\, ,\label{recX1}\\
g^{X}_{k+2}(x) &\!=\!& \left(\strut g^{X}(x) - V'(x)\right) g^{X}_{k}(x) + 
S\Delta_{k}(x)\, ,\label{recX2}\ea
where the $\Delta_{k}(x)$ are degree $d-1$ polynomials that can be 
computed from $g^{X}_{k}$ and that are defined by
\be\label{Delkdef}\Delta_{k}(x) = \Bigl\langle {\tr\over n}\, Y^{k}
{V'(x)-V'(X)\over x-X}\Bigr\rangle\, \cdotp\end{equation}
The relations (\ref{recX1}) and (\ref{recX2}) determine all the 
$g^{X}_{k}$s, and thus $\Gamma(x,y)$, from the single function 
$g^{X}(x)$. Moreover, from (\ref{le2}) we get
\be\label{recX3} g^{X}_{q+2}(x) + g^{X}_{q+2}(-x) = -{\alpha\over 2x}
\left(\strut g^{X}_{q+1}(x) - g^{X}_{q+1}(-x)\right) - \sum_{k+k'=q} 
g^{X}_{k}(x)g^{X}_{k'}(-x)\, .\end{equation}
By using (\ref{recX1}) and (\ref{recX2}) in (\ref{recX3}) we obtain,
for each $q$, equations that close under $g^{X}(x)$ and $g^{X}(-x)$.
Any two such independent equations then determine unambiguously
$g^{X}(x)$. It turns out that the equation for $q=1$ does not yield
anything new, so we use $q=0$ and $q=2$. Eliminating $g^{X}(-x)$ we
get, after a lengthy but straightforward calculation, a closed
equation for $g^{X}(x)$ only. It is a cubic algebraic equation of the
form
\be\label{solXloop} {\cal C}\left(\strut 
x^{2},g^{X}(x)-V'(x)\right) = 0\, ,\end{equation}
where $\cal C$ is defined exactly as in (\ref{ACl}) for $G=-\alpha$.
The deformation polynomials are found explicitly in terms of
correlators to be
\ba \hat\Delta(w) &\!=\!&
2\,\langle{\tr\over n}\, Y\rangle\, ,\label{defXL1}\\
\Delta(x^{2},w) &\!=\!& \Delta_{2}(x) + \Delta_{2}(-x) + xF_{2}(x^{2})
\left(\strut\Delta_{0}(x)-\Delta_{0}(-x)\right)\nonumber\\&& - 
F_{1}(x^{2})\left(\strut \Delta_{0}(x)+\Delta_{0}(-x)\right) + 
{\alpha\over 2x}\left(\strut 
\Delta_{1}(x)-\Delta_{1}(-x)\right)\nonumber\\&& +
\left(\strut \Delta_{0}(x)+\Delta_{0}(-x)\right) w\, .\label{defXL2}\ea
The form of $\Delta$ and $\hat\Delta$ is exactly as discussed at
the end of Section 3.4: the term independent of $w$ in $\Delta$ is a
polynomial of degree $d-1$ in $z=x^{2}$, the term linear in $w$ is a
polynomial of degree $[(d-1)/2]$ in $z$, and $\hat\Delta$ is a
constant. We have thus found a precise match with the result obtained
from the geometric approach. Note that similar cubics were found from
loop equations in \cite{cubiclit} for models that are deformations of 
${\cal N}=2$ gauge theories.\vfill\eject

\subsection{The resolvent for $Y$}

One way to compute the resolvent for $Y$ is to expand the loop
equations at large $x$. From (\ref{le1}) we get a set of linear
equations
\begin{multline}
\sum_{k=0}^{d}t_{k+1}g^{Y}_{k+q}(y) =S\, \langle {\tr\over n}\, 
Y\rangle \delta_{q,0} - {\alpha S\over 2}\,
\langle{\tr\over n} X^{q-1}\rangle\left(\strut 
1-\delta_{q,0}\right) +S y \,\langle{\tr\over n} X^{q}\rangle\\
- y^{2}g^{Y}(y) +
S\!\!\sum_{k+k'=q-1}\langle{\tr\over n} X^{k}\rangle\, 
g^{Y}_{k'}(y)\, ,\label{recY1}\end{multline}
and from (\ref{le2}) we get a set of quadratic equations
\be\label{recY2}
yg^{Y}_{2q-1}(y) = S\,\langle{\tr\over n} X^{2q-1}\rangle - 
{\alpha\over 2} g^{Y}_{2q-2}(y) + {1\over 
2}\sum_{k+k'=2q-2}(-1)^{k}g^{Y}_{k}(y) g^{Y}_{k'}(y)\, .\end{equation}
The $2d$ equations (\ref{recY1}) for $0\leq q\leq d-1$ and
(\ref{recY2}) for $1\leq q\leq d$ close under the $2d$ unknown
$g^{Y}_{k}$ for $0\leq k\leq 2d-1$. We can for example express
linearly the $g^{Y}_{k}$ for $d\leq k\leq 2d -1$ in terms of the
$g^{Y}_{k}$ for $0\leq k\leq d -1$ by using (\ref{recY1}), and then
obtain a set of $d$ quadratic equations for the $d$ unknown
$g^{Y}_{0},\ldots ,g^{Y}_{d-1}$ from (\ref{recY2}). We thus discover
that the resolvent for $Y$ lies at the intersection of $d$ quadrics,
consistently with the result of Section 4.3.2. However, the direct
calculation showing that the set of quadratic equations obtained from
(\ref{recY1}) and (\ref{recY2}) on the one hand and (\ref{expC}) and
(\ref{ACl}) on the other hand are equivalent turns out to be extremely
tedious. 

We are thus going to provide a much simpler proof using the fact that we
have already computed the resolvent for $X$ in Section 5.2.
Let us consider the equation for the unknown $x(y)$
\be\label{lemeq} g^{X}\left(\strut x(y)\right)-V'(x) = y^{2}\, .\end{equation}
We know that $x(y)$ then automatically satisfies (\ref{CL2}). Not all
the solutions to (\ref{CL2}), though, satisfies (\ref{lemeq}). This is
best seen by plugging (\ref{lemeq}) in (\ref{le1}), which yields
\begin{multline}\label{cY} 
\Bigl( yx(y) - {\alpha\over 2}\Bigr) \Bigl( 
y^{2}+V'\left(\strut x(y)\right)\Bigr) + S\langle {\tr\over n}\, 
Y\rangle\\ + S x(y) \Bigl\langle {\tr\over n}\, {1\over 
y-Y}{V'\left(\strut x(y)\right)-V'(X)\over x(y)-X}\Bigr\rangle = 0\, .
\end{multline}
This is a degree $d+1$ algebraic equation for $x(y)$, and thus only
half of the roots of (\ref{CL2}) can actually satisfy (\ref{lemeq}).
By taking the classical limit $S\rightarrow 0$ of (\ref{cY}), we see
that those $d+1$ roots are precisely the roots $x_{i}$ that we have
used in Section 4.3.2 and that were characterized by (\ref{cliLY}). By
looking at the coefficient of $x^{d+1}$ (for the overall 
normalization) and of $x^{d}$ in (\ref{cY}), we then immediately
get the sum $\sigma_{1}=\sum_{i=1}^{d+1}x_{i}(y)$,
\ba\label{solsigma1}\sigma_{1}&\! = \!& -{t_{d}\over 
t_{d+1}}+{\alpha\over 2y} - {g^{Y}(y)\over y}\quad {\rm if}\quad 
d\geq 2\\
&\! = \!& -{t_{1}+y^{2}\over t_{2}}+{\alpha\over 2y} - {g^{Y}(y)\over 
y}\quad {\rm if}\quad d=1\, .\ea
Those equations are equivalent to (\ref{SOLLY1}) and (\ref{SOLLY2}). 
Moreover, we see that the higher symmetric polynomials $\sigma_{k}$, 
$2\leq k\leq d$, are related to the generalized resolvents $g^{Y}_{k}$ 
for $k\leq d-1$. Since the degree $d$ is arbitrary, we conclude that the 
Calabi-Yau geometry actually encodes the full generating function 
$\Gamma$.\goodbreak

\section{Discussion}
\setcounter{equation}{0}

The geometric approach to matrix models is singled out by its
aesthetic features and its relationship with some of the deepest
insights in gauge theory and string theory. It provides an entirely
new perspective on the problem of summing planar diagrams, and
suggests that a whole new class of models could be solved. The
examples that we have studied show that, when appropriate, the
geometric approach, which reduces to the calculation of a blow down 
map, is much more powerful than standard techniques.
There remains, however, many open problems, some of which 
we review below.

\noindent M{\scshape atrix model technology}: it is now clear that all
the irreducible representations of the algebra $\cal A$ of equations
of motion should be taken into account, even though most of the
classic matrix model literature considers only one dimensional
representations, often with the additional one-cut requirement. We
have demonstrated in Section 2 that new qualitative features occur
for higher dimensional representations. It
would be desirable to develop the analytic techniques to deal with the
saddle point equations in those cases, and to have a more general
derivation of the property of eigenvalue entanglement. Another basic
problem is to understand the general conditions under which special
geometry relations can be derived. The geometric approach suggests
that Casimir operators play a special r\^ole in this respect. Another
interesting aspect is that the difficulty of a given model seems to be
directly related to the complexity of the algebra $\cal A$ (the
representation theory and the structure of the center). It is not clear
how this translates in the language of loop equations. An interesting
concrete problem would be understand what makes the algebra and the
loop equations associated with the models (\ref{potential}) special.

\noindent A{\scshape lgebraic geometry}: the infrared slavery of gauge
theories suggests that all the Calabi-Yau geometries of the form
(\ref{2matgeo}), or even (\ref{geogen}) when the corank of the Hessian
of the potential $W$ at critical points is at most two, can be blown
down. It is not clear how this works mathematically. In particular, it
would be desirable to understand the relationship with the results of
\cite{KM} on Gorenstein threefold singularities. A na\"\i ve guess
would have been that blow down can only be found for the cases
constructed in \cite{CKV}, but clearly the models (\ref{potential})
are much more general, and already the model (\ref{lll}) that we have
solved provides a counter-example. A startling fact is that {\it the
calculation of the blow down map $\pi$ is essentially equivalent to
finding the solution of the associated matrix model}. It might be
possible to devise an algorithm that computes $\pi$. Another
interesting feature is the interplay between the non-commutative
structure and the singularity structure of the blown down geometry,
which must reproduce the representation theory of the model. In some
sense the blow down map knows about D-branes.

\noindent G{\scshape eometric approach technology}: we have explained in 
Section 4 how to calculate the resolvents from the geometry. It is not 
clear to what extent this approach can be applied in general. The 
issue is to understand the fibered structure of the geometries, which 
is probably related to the structure of the center of $\cal A$. We 
would also like to understand what are the most general 
matrix model correlators encoded in the geometry, and also how to 
extract the non-planar contributions. The answer can probably be found 
in the topological string set-up \cite{top}.

\noindent D{\scshape -branes on Calabi-Yau}: an interesting question
is to ask what is the most general matrix model that can be engineered
by putting branes in a Calabi-Yau. This is the ``reverse geometric
engineering'' problem \cite{beren}. When there is no moduli space,
which is the generic case we have been considering, we have very
little insight into the solution of that problem. We have been able to
construct in Section 3 a large class of models, but it might be
possible to find additional theories by considering more general
perturbations to the geometry (\ref{geo1}). For example, one would
like to know, given a $\pone$ in a Calabi-Yau, what is the most
general geometry for which the obstruction to the versal deformation
space of the $\pone$ is integrable in terms of a potential, taking
into account the non-commutativity of the variables. The solution is
probably most naturally expressed in terms of some ``regularity''
conditions on the algebra $\cal A$. Even if we could understand the
reverse geometric engineering, there would remain the fundamental
question of why and when the geometric transition conjecture is valid.
The phenomenon it describes is very reminiscent, both physically and
mathematically, to the ``continuation to negative radius'' found in
two dimensional $\sigma$ models (see for example \cite{CecV}). In this
latter case, we know that some non-geometrical phases are possible.
This suggests that we may presently only catch a glimpse of the full
story for what can happen beyond the singular (infinite gauge
coupling) point. The matrix models provide in principle a powerful
tool to study that question. A full understanding would amounts to
describing the space of vacua in string theory.

\noindent ${\cal N}=1$ {\scshape gauge theories}: we have used
throughout the full non-perturbative Dijkgraaf-Vafa conjecture. This
includes, consistently with special geometry, keeping the
Veneziano-Yankielowicz term and the higher powers in the glueball
superfields $S_{i}$, even when the latter are perturbatively zero in
the chiral ring. There is no proof of this conjecture at the moment.
When a direct solution of the matrix model exists, as described in
Section 5, our result can be interpreted as providing a non-trivial
check of the consistency betweeen the Dijkgraaf-Vafa conjecture and
the geometric transition picture. The two-matrix models we have
studied are particularly interesting because they are not deformation
of ${\cal N}=2$ theories (which requires either one or three adjoint
fields), and thus there is no Seiberg-Witten curve. The matrix model
is then the only tool at our disposal. It would be very interesting to
work out the quantum space of parameters for these theories, along the
lines of \cite{fer1,fer2}. In particular, it is in principle possible
to study quantitatively the phase transition between the Higgs and
confining phases on parameter space \cite{S}.

As a final comment, we would like to note that we have put the
emphasis on matrix models and thus on theories with only adjoint
fields. However, all the questions we have addressed are also relevant
to the more general set-up of quiver theories.

\subsection*{Acknowledgements}

I would like to thank D.~Berenstein, S.~Katz, W.~Lerche, C.~Vafa and
particularly R.~Dijkgraaf for very useful discussions, as well as the
organizers of the Amsterdam summer workshop ``String Theory and Quantum
Gravity,'' June 2003, where some of the results of this article were
first presented. This work was supported in part by the Swiss National
Science Foundation. The author is on leave of absence from Centre
National de la Recherche Scientifique, Laboratoire de Physique
Th\'eorique de l'\'Ecole Normale Sup\'erieure, Paris,
France.
\vfill\eject

\renewcommand{\thesection}{\Alph{section}}
\renewcommand{\thesubsection}{\arabic{subsection}}
\setcounter{section}{0}
\section{Saddle point equations and special geometry}
\setcounter{equation}{0}
\subsection{Continuum limit of singular sums}

Let us consider the sum
\be\sigma_{n} (x) = {1\over n}\sum_{i=1}^{n}{1\over 
x-x_{i}}\,\cdotp\end{equation}
We assume that when $n\rightarrow\infty$, the distribution of 
eigenvalues $(x_{i})_{1\leq i\leq n}$ goes to a smooth function $\rho$
with compact support, and we want to compute
\be\label{defss}\sigma (x) = \lim_{n\rightarrow\infty}\sigma_{n}(x)\, 
.\end{equation}
If $x\not\in {\rm Support}[\rho]$, we have
\be\label{limit}\sigma(x)=g(x)\, ,\end{equation}
where the analytic function $g(x)$ is defined by
\be\label{sadef}
g(x)=\int_{-\infty}^{+\infty} {\rho(z)\,\d z\over x-z}\,
\cdotp\end{equation}
If $x\in {\rm Support}[\rho]$, $g(x)$ is ambiguous because
$g$ has a branch cut. Let us label the eigenvalues in such a way
that $x_{1}<x_{2}<\cdots <x_{n}$, and pick $x_{j}<x<x_{j+1}$ with
$\delta^{+}_{n}=x_{j+1}-x$ and $\delta^{-}_{n}=x-x_{j}$. Let us define
\be\label{deldef}{\delta^{+} (x)\over\delta^{-} (x)}=
\lim_{n\rightarrow\infty} 
{\delta^{+}_{n}\over\delta^{-}_{n}}\,\cdotp\end{equation}
We can then compute
\ba\sigma(x)&\!=\!& \lim_{n\rightarrow\infty}
\left(\int_{-\infty}^{x-\delta^{-}_{n}} + \int_{x
+\delta^{+}_{n}}^{+\infty}\right) {\rho(z)\,\d z\over x-z}\nonumber\\
&\! =\! & {1\over 2} \left( \strut
g(x+i\epsilon)+g(x-i\epsilon)\right) + \rho(x) \ln
{\delta^{+} (x)\over\delta^{-} (x)}\,\cdotp\label{limitp}\ea
One can use the above formula to deduce equation (\ref{spnc})
from (\ref{sadL}). The same reasoning also yields 
the standard result
\be\label{standardlim} \lim_{n\rightarrow\infty}\left( {1\over 
n}\sum_{j\not = i}{1\over x_{i}-x_{j}}\right) = 
{1\over 2} \left( 
\strut g(x_{i}+i\epsilon)+g(x_{i}-i\epsilon)\right)\end{equation}
because at large $n$ we have in that case $\delta^{+}=x_{i+1}-x_{i} =
1/(n\rho(x_{i})) = x_{i}-x_{i-1}=\delta^{-}$.

\subsection{Analysis of the saddle point equations}

Let us discuss briefly the saddle point equation 
(\ref{spnc}) supplemented with the condition (\ref{evenrho}). 
If the intervals $I^{X}_{K}$ are filled with eigenvalues in one 
dimensional representations, we have
\be\label{spdL1} g^{X}(x + i\epsilon) + g^{X}(x-i\epsilon) + 
g^{X}(-x) = V'(x) + {\alpha^{2}\over 4 x^{2}}\quad {\rm for}\ x\in 
I^{X}_{K}\, .\end{equation}
Moreover, by taking the even and odd parts of (\ref{spnc}) for two 
dimensional representations, and using (\ref{evenrho}) and 
(\ref{densfor}), we get
\ba &&\hskip -1.5cm
S\rho^{X}(-x)\ln{\delta^{+}(x)\over\delta_{-}(x)} ={\alpha^{2}\over 4 
x^{2}} -  F_{1}(x^{2}) - 3x F_{2}(x^{2}) \nonumber\\ &&\hskip 3.5cm
-{3\over 2}\left(\strut
g^{X}(x+i\epsilon) + g^{X}(x-i\epsilon)\right) ,
\label{spdL2}\\ &&\hskip -1.5cm
g^{X}(x+i\epsilon) = g^{X}(-x + i\epsilon) - 2xF_{2}(x^{2})\, ,\quad 
{\rm for}\ x\in \tilde I^{X}_{K}\cup\tilde {I'}^{X}_{K}\, .\label{spdL3}
\ea
Let us denote $g^{X}_{\rm first}(x) = g^{X}(x)$ on the physical sheet.
The equation (\ref{spdL1}) shows that the cuts $I^{X}_{K}$ glue the 
physical sheet with a second sheet where the solution is 
\be\label{2nd}
g^{X}_{\rm second}(x) =-g^{X}_{\rm first}(x) - 
g^{X}_{\rm first}(-x) + V'(x) + {\alpha^{2}\over 4x^{2}}\, \cdotp\end{equation}
Due to the term $g^{X}_{\rm first}(-x)$, this second sheet contains
new cuts $I^{'X}_{K}$ that are the image with respect to $x=0$ of the
physical cuts $I^{X}_{K}$. Using again (\ref{spdL1}), we see that the
new cuts glue the second sheet with a third sheet with
\ba g^{X}_{\rm third}(x)&\! =\! & - g^{X}_{\rm first}(x)
\nonumber\\&&
-\Bigl(-g^{X}_{\rm first}(-x)-g^{X}_{\rm first}(x) +
V'(-x)+{\alpha^{2}\over 4 x^{2}}\Bigr) + 
V'(x) + {\alpha^{2}\over 4 
x^{2}}\nonumber\\&\! =\! & g^{X}_{\rm first}(-x) - 2xF_{2}(x^{2})\, 
.\label{third}\ea
Equation (\ref{spdL3}) then shows that this third sheet is glued to
the first sheet by the cuts $\tilde I^{X}_{K}$ and $\tilde
I^{'X}_{K}$. Overall $g^{X}$ has generically a three-sheeted
structure, except when only two dimensional representations are
present in which case the second sheet is absent. The general form of
the cubic equation satisfied by $g^{X}$ is derived both in Section 4
and in Section 5. A special form of this cubic in the one-cut
assumption and for $\alpha =0$ was found in \cite{EZJ}. Let us also
note that models where a similar cubic appears have been studied
recently in the literature \cite{cubiclit}.

\subsection{The effective potential for $Y$}

We have argued in Section 2.2 that the singularity structure found 
for the background force (\ref{laubf}) was generic. We want to illustrate 
this fact by studying the effective potential for $Y$ in the model 
(\ref{ldef}), defined by
\be\label{yeffp1}e^{-{n^{2}\over S} V_{{\rm eff}}(Y)} = 
e^{-{n\alpha\over S}\tr Y}\int\d X
e^{-{n\over S}\tr (XY^{2}+V(X))}\, .\end{equation}
We know that $V_{\rm eff}$ must have singularities when
$\deg V\geq 2$ to make the associated saddle point equations consistent 
with two dimensional representations.
This is a little bit puzzling at first, because the effective 
potential for $Y$ is essentially the same as the effective potential 
for the ordinary two-matrix model discussed in Section 2.2.2, except 
that it is a function of the square of the matrix instead of the 
matrix itself. A simple way to understand what is going on is to use 
the Itzykson-Zuber formula \cite{IZint} to derive
\ba  && \hskip -1cm e^{-{n^{2}\over S} V_{{\rm eff}}(Y)} =
{e^{-{n\alpha\over S}\sum_{i} y_{i}}
\over\prod_{i<j}\left(\strut(y_{i}-y_{j})(y_{i}+y_{j})\right)}
\nonumber\\ &&\hskip 2cm 
\int\prod_{i}\d x_{i}\, \prod_{i<j}(x_{i}-x_{j})\,
e^{-{n\over S}\sum_{i}(x_{i}y_{i}^{2} + V(x_{i}))}.\label{effyIZ}\ea
This formula shows that the only singularities may be at 
$y_{i}=y_{j}$ or $y_{i}=-y_{j}$. To find which singularities 
actually occur, we can evaluate (\ref{effyIZ}) in the limit 
$S\rightarrow 0$. This yields 
\be\label{effyIZcl} e^{-{n^{2}\over S} V_{{\rm eff}}(Y)} 
\mathrel{\mathop{\kern 0pt \propto}\limits_{S\rightarrow 0}^{}}
{\prod_{i<j}\left(\strut\chi(y_{i}^{2})-\chi(y_{j}^{2})\right)
\over\prod_{i<j}\left(\strut(y_{i}-y_{j})(y_{i}+y_{j})\right)}\,\cvp\end{equation}
where the function $\chi (y^{2})={V'}^{-1}(-y^{2})$ minimizes $\chi
y^{2} + V(\chi)$. When $y_{i}\rightarrow y_{j}$, the pole in the
denominator of (\ref{effyIZcl}) is cancelled by a zero of the
denominator, consistently with the fact that no singularities are
expected at $y_{i}=y_{j}$. On the other hand, because $\chi$ is
$d$-sheeted, if we analytically continue $y_{i}$ from $y_{i}=y_{j}$ to
$y_{i}=-y_{j}$ we can go to a different sheet
$\chi(y_{i}^{2})\rightarrow\tilde\chi (y_{i}^{2})$ and thus produce a
pole in (\ref{effyIZcl}). We read from (\ref{effyIZ}) that the
relevant singular part in $V_{\rm eff}(Y)$ is exactly the same as that
for $V_{\rm eff}(X)$ in (\ref{veffX}). In particular the eigenvalues
will be entangled for two dimensional representations, implying that
\be\label{entY}\rho^{Y}(y)=\rho^{Y}(-y)\quad {\rm for}\ y\in \tilde 
I^{Y}_{K}\, .\end{equation}
\subsection{Derivation of special geometry relations}

An elegant way to derive special geometry relations is to set up a 
variational formulation of the saddle point equations. Let us first 
deal with the description in terms of the matrix $X$.
One must be careful because
\be\label{naivew} n {\partial V_{\rm eff}\over\partial 
x_{i}}\lower 3pt\hbox{$\Bigr|_{x_{i}=x}$}\not = {\d\over\d x}{\delta 
V_{\rm eff}(\rho^{X})\over\delta\rho^{X}(x)}\end{equation}
due to the logarithmic term in (\ref{limitp}). Let us consider the 
functional
\ba &&\hskip -.55cm \F = -S V_{\rm eff}(\rho^{X}) + S^{2}\int\!\d 
x\d\tilde x\, \rho^{X}(x)\rho^{X}(\tilde x)\ln |x-\tilde x| \nonumber\\
&&\hskip .3cm
+\sum_{K=1}^{d+2} \ell_{K}\Bigl( \int_{I^{X}_{K}}S\rho^{X}(x)\,\d x - 
S_{K}\Bigr) + \sum_{K=1}^{[(d-1)/2]}\tilde\ell_{K}
\Bigl( \int_{\tilde I^{X}_{K}}S\rho^{X}(x)\,\d x - 
\tilde S_{K} \Bigr) \nonumber\\
&& \hskip .3cm +\sum_{K=1}^{[(d-1)/2]}\int_{\tilde I^{X}_{K}} S
\left(\rho^{X}(x) - \rho^{X}(-x)\right) L_{K}(x)\, \d x\, .\label{Fvar}
\ea
By varying $\F$ with respect to the Lagrange multipliers $\ell_{K}$
and $\tilde\ell_{K}$ we obtain
\be\label{Lag1c} \int_{I^{X}_{K}}S\rho^{X}\d x = {1\over 
2i\pi}\oint_{\alpha^{X}_{K}}g^{X} \d x = S_{K}\, ,\quad
\int_{\tilde I^{X}_{K}}S\rho^{X}\d x = {1\over 
2i\pi}\oint_{\tilde\alpha^{X}_{K}}g^{X} \d x = \tilde S_{K}\, .
\end{equation}
The notation for the contours is similar to that of Figure 1. The
equations (\ref{Lag1c}) implement the relations (\ref{Skdef}), with
$\tilde S_{K}$ associated with two dimensional representations. By
varying $\F$ with respect to the Lagrange multipliers $L_{K}(x)$, we
get the constraint (\ref{evenrho}). Finally, by varying $\F$ with
respect to $\rho^{X}(x)$ we get the saddle point equations
\ba &&\hskip -1.5cm 0= -{\delta V_{\rm eff}\over\delta\rho^{X}(x)} + 2 
S\int\!\d\tilde x\,\rho^{X}(\tilde x)\ln |x-\tilde x| + \ell_{K}
\quad {\rm for}\ x\in I^{X}_{K}\, ,\label{spL1}\\
&&\hskip -1.5cm 0= -{\delta V_{\rm eff}\over\delta\rho^{X}(x)} +L_{K}(x)+ 2 
S\int\!\d\tilde x\,\rho^{X}(\tilde x)\ln |x-\tilde x| + \tilde\ell_{K} 
\quad  \!\!{\rm for}\ x\in \tilde I^{X}_{K},\label{spL2}\\
&&\hskip -1.5cm 
0= -{\delta V_{\rm eff}\over\delta\rho^{X}(x)} -L_{K}(-x)+ 2 
S\int\!\d\tilde x\,\rho^{X}(\tilde x)\ln |x-\tilde x|
\quad {\rm for}\ x\in \tilde I^{'X}_{K}\, .\label{spL3}\ea
One can check that the derivatives of the above equations are
equivalent to (\ref{spdL1}), (\ref{spdL2}) and (\ref{spdL3}), with the
identification
\be\label{LKvsab} L'_{K}(x) = 
S\rho^{X}(-x)\ln{\delta^{+}(x)\over\delta^{-}(x)}
\quad {\rm for}\ x\in\tilde I^{X}_{K}\, .\end{equation}
Moreover, the partial derivatives 
with respect to $S_{K}$ and $\tilde S_{K}$ take very simple forms, 
\be\label{ddsp} {\partial\F\over\partial S_{K}} = -\ell_{K}\, ,\quad 
{\partial\F\over\partial \tilde S_{K}} = -\tilde\ell_{K}\,\cdotp\end{equation}

Using (\ref{veffX}) and (\ref{spL1}), it is straightforward to compute
\ba\oint_{\gamma^{X}_{K}}g^{X}(x)\,\d x &\!=\!& 
\int_{\mu_{0}}^{a_{K}}\Bigl( 2g^{X}(x) - {\d\over\d x}{\delta V_{\rm 
eff}\over\delta\rho^{X}(x)}\Bigr)\nonumber\\
&\!=\!& -\ell_{K} + {\delta V_{\rm 
eff}\over\delta\rho^{X}(\mu_{0})} - 2 S \ln\mu_{0} + {\cal 
O}(1/\mu_{0})\, ,\label{conti1} \ea
which yields explicitly
\be\label{res1}{\partial\F\over\partial S_{K}} = 
\lim_{\mu_{0}\rightarrow\infty}\Bigl( \oint_{\gamma^{X}_{K}}g^{X}\d 
x - V(\mu_{0}) + S\ln\mu_{0}\Bigr) .\end{equation}
Similarly, using (\ref{spL2}), (\ref{spL3}) and (\ref{evenrho}) we get
\be\label{conti2}
\oint_{\tilde\gamma^{X}_{K}}g^{X}(x)\, \d x =-\tilde\ell_{K}+ 
{\delta V_{\rm eff}\over\delta\rho^{X}(\mu_{0})}+ {\delta V_{\rm 
eff}\over\delta\rho^{X}(-\mu_{0})}- 4 S \ln\mu_{0} + {\cal 
O}(1/\mu_{0})\, , \end{equation}
yielding
\be\label{res2}
{\partial\F\over\partial \tilde S_{K}}=
\lim_{\mu_{0}\rightarrow\infty}\Bigl(
\oint_{\tilde\gamma^{X}_{K}}g^{X}\d x -
V(\mu_{0})-V(-\mu_{0}) + 2S\ln\mu_{0}\Bigr) .\end{equation}

Using the results of Section A.3, we can repeat the discussion above 
by replacing the matrix $X$ by the matrix 
$Y$. There is an additional $K$-independent term
\be\label{addterm} \int {\delta V_{\rm eff}(\rho^{Y})
\over\delta\rho^{Y}(y)}\, \rho^{Y}(y)\, \d y - {\partial\over\partial 
S}\left(\strut S V_{\rm eff}\right)\end{equation}
in (\ref{ddsp}) because $V_{\rm eff}(\rho^{Y})$ is not linear in 
$\rho^{Y}$. This term does not affect the relations involving compact 
cycles,
\be\label{Ysg}
{\partial\F\over\partial S_{K}}
- {\partial\F\over\partial S_{K'}} = 
\oint_{\beta^{Y}_{K,K'}}g^{Y}\d y\, ,\quad
{\partial\F\over\partial \tilde S_{K}}
- {\partial\F\over\partial 
\tilde S_{K'}} = \oint_{\tilde\beta^{Y}_{K,K'}}g^{Y}\d y\, .\end{equation}
\section{RG flow theorem for normal bundles}
\setcounter{equation}{0}

In this Appendix, we provide a proof of the conjecture in Section 3.2.1 
in the special case $n=1$ relevant to two-matrix models. The proof is 
a direct consequence of two lemmas.

\noindent L{\scshape emma} 1: Let $\pone$ in the geometry
(\ref{2matgeo}) at a given critical point $(x,y)$ of the
superpotential (\ref{genW2}). Then the transition function
characterizing the normal bundle of the $\pone$ can be cast in the
form
\be\label{NT} T(z)=\begin{pmatrix}z^{-1} & 0\\
\partial^{2}_{y}W + z \partial_{x}\partial_{y}W
+ z^{2}\partial^{2}_{x}W\ & z^{3}\end{pmatrix} .\end{equation}

\noindent \noindent P{\scshape roof:} To calculate the transition
function, we expand $w_{1}=w_{1}(z) + \delta_{1}$,
$w_{2}=w_{2}(z)+\delta_{2}$, $w'_{1}=w'_{1}(z)+\delta'_{1}$ and
$w'_{2}=w'_{2}(z)+\delta'_{2}$ in (\ref{2matgeo}), keeping only the
linear terms in the $\delta$s. By redefining
\be\begin{split} \delta_{2}&\rightarrow
\delta_{2}-\delta_{1}\sum_{i\geq 3}z^{i-3}\sum_{j\geq 0}
E_{i-j}^{(1+j)}(x)y^{j}/j!\, ,\\
\delta'_{2}&\rightarrow
\delta'_{2}+\delta'_{1}\sum_{i>0}{z'}^{i-1}\sum_{j\geq 0}
E_{-i-j}^{(1+j)}(x)y^{j}/j!\, ,\end{split}\end{equation}
we obtain
\be\begin{pmatrix}\delta'_{1}\\ \delta'_{2}\end{pmatrix} = T(z)
\begin{pmatrix} \delta_{1}\\ \delta_{2}\end{pmatrix} .\end{equation}

\noindent L{\scshape emma} 2: Let $\cal E$ be a holomorphic vector 
bundle over $\pone$ with structure group ${\rm GL}(2,\mathbb C)$ and 
transition function
\be T(z) = \begin{pmatrix} z^{-1} & 0\\ a + bz + cz^{2} & 
z^{3}\end{pmatrix}.\end{equation}
Let $r$ be the corank of the quadratic form $Q=\bigl(\begin{smallmatrix}
a&b\\ b&c\end{smallmatrix}\bigr)$. Then ${\cal E}=\O(r-1)\oplus\O(-r-1)$.

\noindent P{\scshape roof}: We want to construct gauge 
transformations $P_{\rm N}(z)$ and $P_{\rm S}(z'=1/z)$
such that
\be P_{\rm S}TP_{\rm N}^{-1} = \begin{pmatrix}
z^{1-r}&0\\ 0&z^{1+r}\end{pmatrix}.\end{equation}
The important property is that $P_{\rm N}(z)$, $P_{\rm N}^{-1}(z)$,
$P_{\rm S}(z')$ and $P_{\rm S}^{-1}(z')$ must be holomorphic 
functions. It is not difficult to check that the following matrices 
have the required properties: for $ac-b^{2}\not = 0$ and $b\not = 0$,
\be P_{\rm N} = \begin{pmatrix} ac/b - b - cz& (a/b - z)z\\
c^{2}&c z-b\end{pmatrix}\, ,\
P_{\rm S}=\begin{pmatrix}-a^{2}/b& az'/b-1\\
b^{2}-ac+abz' & (c-bz')z'\end{pmatrix} ,\end{equation}
for $b=0$, $a\not =0$ and $c\not =0$,
\be\begin{split} P_{\rm N} &= \begin{pmatrix} -c(1+z)/a & 1/c - z(z+1)/a\\
c&z \end{pmatrix}\, ,\\
P_{\rm S}&=\begin{pmatrix} 1-az'/c&{z'}^{2}/c -(z'+1)/a\\ -a&z'
\end{pmatrix},\end{split}\end{equation}
for $a=\alpha^{2}$, $b=\alpha\beta$, $c=\beta^{2}$, $\alpha\not=0$ 
and $\beta\not =0$,
\be P_{\rm N} = \begin{pmatrix}\beta^{2}&z\\ -\beta^{3}/\alpha& 1-\beta 
z/\alpha \end{pmatrix}\, ,\
P_{\rm S}=\begin{pmatrix} -\alpha^{2}z'-\alpha\beta &{z'}^{2}\\
-\alpha^{2} & z'-\beta/\alpha\end{pmatrix} ,\end{equation}
for $a=b=0$ and $c\not =0$,
\be P_{\rm N} = \begin{pmatrix} 0&-1/c\\ c&z\end{pmatrix}\, ,\
P_{\rm S}=\begin{pmatrix} 1&-{z'}^{3}/c\\ 0&1\end{pmatrix} ,\end{equation}
for $b=c=0$ and $a\not =0$,
\be P_{\rm N} = \begin{pmatrix} a&z^{3}\\ 0&-1/a\end{pmatrix}\, ,\
P_{\rm S}=\begin{pmatrix} 0&1\\ 1&-z'/a\end{pmatrix} ,\end{equation}
and finally for $a=b=c=0$, $P_{\rm N}=P_{\rm S}=I$.

\end{document}